\documentclass[aps,pra,preprint,showpacs,superscriptaddress,longbibliography]{revtex4-1}
\usepackage{amsmath,amssymb,mathtools,bm,mathrsfs,graphicx, braket,amsthm,enumerate,times}
\usepackage[colorlinks=true,citecolor=blue,linkcolor=blue]{hyperref}
\usepackage{longtable}
\usepackage{multirow} 

\newcommand{\calC}{\mathcal{C}}
\newcommand{\calH}{\mathcal{H}}
\newcommand{\calL}{\mathcal{L}}
\newcommand{\hBdG}{h_{\mbox{\scriptsize BdG}}}
\newcommand{\thBdG}{\tilde{h}_{\mbox{\scriptsize BdG}}}
\newcommand{\tthBdG}{\tilde{\tilde{h}}_{\mbox{\scriptsize BdG}}}

\newcommand{\tXi}{\tilde{\Xi}}

\newcommand{\Ztwo}{$\mathbb{Z}_2$}

\newcommand{\frakb}{\mathfrak{b}}

\newcommand{\Bphi}{\boldsymbol{\phi}}
\newcommand{\tBphi}{\tilde{\boldsymbol{\phi}}}
\newcommand{\ttBphi}{\tilde{\tilde{\boldsymbol{\phi}}}}
\newcommand{\tphi}{\tilde{\phi}}
\newcommand{\ttphi}{\tilde{\tilde{\phi}}}

\newcommand{\Bpsi}{\boldsymbol{\psi}}

\newcommand{\sgn}{\mathrm{sgn}}
\newcommand{\epsmin}{\mathcal{E}_{\mbox{\scriptsize min}}}

\usepackage{etoolbox}

\makeatletter
\patchcmd{\frontmatter@abstract@produce}
  {\vskip200\p@\@plus1fil
   \penalty-200\relax
   \vskip-200\p@\@plus-1fil}
  {}
  {}
  {}
\def\frontmatter@preabstractspace{3cm}
\makeatother

\begin{document}

\title{
Universal finite-size gap scaling of the quantum Ising chain}

\author{Masaki Oshikawa}

\email{oshikawa@issp.u-tokyo.ac.jp}
\affiliation{Institute for Solid State Physics, University of Tokyo, Kashiwa 277-8581, Japan} 
\affiliation{Kavli Institute for the Physics and Mathematics of the Universe (WPI), University of Tokyo, Kashiwa 277-8583, Japan}
\affiliation{Kavli Institute for Theoretical Physics, University of California Santa Barbara, CA 93106, USA}

\date{\today}

\begin{abstract}
I study the universal finite-size scaling function for the lowest
gap of the quantum Ising chain with a one-parameter family
of ``defect'' boundary conditions, which includes periodic,
open, and antiperiodic boundary conditions as special cases.
The universal behavior can be described by the Majorana fermion
field theory in $1+1$ dimensions, with the mass proportional to
the deviation from the critical point.
Although the field theory appears to be symmetric with respect
to the inversion of the mass (Kramers-Wannier duality),
the actual gap is asymmetric, reflecting the spontaneous symmetry
breaking in the ordered phase which leads to the
two-fold ground-state degeneracy in the thermodynamic limit.
The asymptotic ground-state degeneracy in the ordered phase
is realized by (i) formation of a bound state at the defect
(except for the periodic/antiperiodic boundary condition) and (ii)
effective reversal of the fermion number parity in one
of the sectors (except for the open boundary condition),
resulting in a rather nontrivial crossover ``phase'' diagram
in the space of the boundary condition (defect strength) and
mass.
\end{abstract} 

\maketitle

\clearpage

\section{Introduction}
\label{sec.Introduction}

Universality in critical phenomena is one of the most important
concepts in statistical physics.
At the critical point, the long-distance properties are
governed by a fixed point in terms of
Renormalization Group (RG), which is nothing but a scale-invariant
(conformal) field theory.
This is best established in 2 dimensions, or for ``relativistic''
quantum critical point in 1 spatial dimension.
The corresponding field theories are conformal field theories (CFTs)
in 1+1 dimensions~\cite{CFT-YellowPages},
which are governed by the infinite-dimensional
Virasoro algebra. This correspondence has led to many quantitative
predictions which have been confirmed numerically and experimentally.
In particular, the finite-size scaling of energy gaps is important.
In the thermodynamic (infinite-size) limit of a critical system,
the energy spectrum is continuous above the ground state.
However, in a finite-size system, the energy spectrum is generally
discrete, and there is a nonvanishing gap between energy eigenvalues.
Understanding this spectrum is of practical importance, since
only finite-size systems are accessible by most of numerical calculations.
In particular, the exact diagonalization of a finite system provides
a very precise energy spectrum.
According to the CFT in 1+1 dimensions, the excitation energies $\Delta E_n$
(energy eigenvalues relative to the ground state) of a finite-size
system of length $L$ with the periodic boundary condition
are related to the scaling dimensions $x_n$ as~\cite{Cardy-CFTFSS_1984}
\begin{equation}
 \Delta E_n = \frac{2\pi x_n}{L},
\end{equation}
thanks to the conformal mapping between the infinite plane and
the infinite cylinder.
The ground-state energy also obeys the universal correction
\begin{equation}
 E_0 = \varepsilon_0 L - \frac{\pi c}{6 L} + O(\frac{1}{L^2}),
\end{equation}
where $c$ is the central charge which characterizes the CFT,
and $\varepsilon_0$ is the non-universal ground-state energy density
in the thermodynamic limit.
Similar relations for the open boundary condition are also
obtained~\cite{Cardy-EffectBC_1986}.
These celebrated relations have been successfully tested and
utilized for many quantum critical systems in 1+1 dimensions.
While the analysis becomes more involved in higher dimensions,
it is recently extended to 2+1 dimensions~\cite{CFTFSS-highD_2016}.

Even away from the critical point, in the vicinity of the critical point
the long-distance properties are governed by a universal RG flow
away from the fixed point, which corresponds to a massive field theory.
This should again determine universal properties,
including the scaling of the finite-size energy spectrum.
In field theory, the finite-size correction to the excitation gap in massive
field theories has been known as L\"{u}scher's
formula~\cite{Luescher1986-I,Luescher1986-II,KlassenMelzer1991},
which is useful to extract scattering amplitudes from numerical simulations.
Although L\"{u}scher's formula works in any dimensions,
its structure can be elucidated in
more details in 1+1 dimensions~\cite{Bajnok-FTLuescher_2018}.
The perturbation theory in terms of the ``mass'' (gap-opening)
term added to the CFT in 1+1 dimensions is also very useful
for an accurate determination of quantum critical points of
quantum lattice models in 1 dimension, and was dubbed as
``level spectroscopy''~\cite{OkamotoNomura-LS_1992,Okamoto-LS-Review_2002}.
The finite-size scaling of the energy spectrum has been also
discussed in the context of integrable models, such as $S=1/2$
XXZ and XYZ chains.
Their scaling limits provide lattice regularizations of integrable field
theory in $1+1$ dimensions, such as sine-Gordon and Thirring models.
Thus the universal finite-size gap scaling function can be obtained
in these cases~\cite{
DDV-FT-1995,
Fioravanti-etal-DDV-exsG,
FeveratiRavaniniTakacs-odd-sG,
FeveratiRavaniniTakacs-sG-NLIE,
Bajnok-etal-k-folded-sG}.

However, it appears that, some interesting structures
in the universal finite-size scaling of the gap, especially
in connection to quantum phase transitions involving spontaneous
symmetry breakings,
have not been appreciated in full details.
In this paper, I will discuss the universal finite-size scaling
of the lowest energy gap (the gap between the ground state and
the first excited state in a finite size system)
in the quantum (transverse-field) Ising chain away from the critical point
but in the scaling limit.
The quantum Ising chain, defined by the Hamiltonian
\begin{equation}
\calH_I = - J \sum_j  \sigma^z_j \sigma^z_{j+1}
- \Gamma \sum_j \sigma^x_j ,
\label{eq.qIsingH}
\end{equation}
is a prototypical model of quantum phase transitions and quantum
critical phenomena~\cite{Sachdev-QPTBook}.
Without losing generality, I assume $J>0$.
The first term is
the standard Ising interaction, which is identical
to that in the classical Ising chain, and tend to align the spins
in the same direction ($\sigma^z=\pm 1$) in the ground state.
The second, transverse field term, flips $\sigma^z$ and thus
tend to disorder the spins in $\sigma^z$-basis.
As a result of competition between these two terms,
a quantum phase transition between the ordered phase $\Gamma < \Gamma_c$
and the disordered phase $\Gamma > \Gamma_c$ occurs at the
critical point $\Gamma = \Gamma_c$.
From the exact solution, which will be discussed below
we know that the quantum phase transition
is of second order, and that $\Gamma_c = J$.

The Ising chain is exactly solved~\cite{Pfeuty_1970,Kogut-RMP1979}
by the Jordan-Wigner transformation~\cite{Jordan-Wigner,LSM}
to the free fermion Hamiltonian
\begin{equation}
\calH_K = 
- t \sum_j \left( c^\dagger_j c_{j+1} + c^\dagger_{j+1} c_j \right)
- \mu \sum_j c^\dagger_j c_j
+ \sum_j \left( \Delta c_j c_{j+1} + \Delta^* c^\dagger_{j+1} c^\dagger_j \right),
\label{eq.Kitaevchain}
\end{equation}
where
\begin{align}
 t = \Delta = & J \in \mathbb{R}, \\
 \mu = 2 \Gamma .
\end{align}
While the Hamiltonian~\eqref{eq.Kitaevchain}
contains the pair creation/annihilation terms,
they can be eliminated by a canonical (Bogoliubov) transformation
so that the resulting Hamiltonian is completely diagonalized.
Physically, the pair creation/annihilation terms
in Eq. may be realized by a proximity
effect to a bulk superconductor.
Thus Eq.~\eqref{eq.Kitaevchain} can represent (non-interacting)
electrons in one-dimensional channel attached to a superconductor.
It is nothing but the ``Kitaev chain'' Hamiltonian~\cite{Kitaev-chain},
which exhibits
Majorana zero modes localized at the open ends, in the topological phase.
The Kitaev chain (and its possible experimental realizations) is studied
vigorously in recent years, in relation to topological quantum computation.
The Kitaev chain~\eqref{eq.Kitaevchain} has a second-order quantum
phase transition when
\begin{equation}
 |\mu| = | 2 t |, 
\end{equation}
which corresponds to $|\Gamma|= |J|$ in the Ising chain.
In the case of the Kitaev chain, it is a quantum phase transition
between the topological (with Majorana zero modes at the ends) and
non-topological phases.

While the Ising chain and Kitaev chain are ``equivalent'' since
they are mapped to each other by the Jordan-Wigner transformation,
the physical (local) operator content is different.
The non-locality of the Jordan-Wigner transformation also introduces
subtleties in the boundary condition.
The periodic boundary condition for the Ising chain is not translated
simply to the periodic boundary condition for the Kitaev chain.
These issues have been also well understood in terms of the CFT
describing the quantum critical point.
As discussed in detail in Appendix~\ref{app.Ising_cont},
the continuum limit of the Kitaev chain near the critical point
is the Majorana fermion field theory in 1+1 dimensions,
defined by the Lagrangian density
\begin{equation}
\calL_M = \bar{\Bpsi} (i \partial_\mu \gamma^\mu - m ) \Bpsi
\end{equation}
with the two-component real field $\Bpsi$.
The corresponding Hamiltonian in the chiral basis is
\begin{equation}
 \calH_I \sim \frac{1}{2} \int dx\; \Bpsi^T(x) \hBdG(x) \Bpsi(x),
\end{equation}
where
\begin{equation}
 \Bpsi(x) =
\begin{pmatrix}
 \psi_R(x)
\\
 \psi_L(x)
\end{pmatrix}
\end{equation}
is the real field $\psi_{R,L}^\dagger(x) = \psi_{R,L}(x)$ and
\begin{equation}
 \hBdG = -i \partial_x \tau^z - m \tau^y .
\end{equation}

General quantum states in this system can be constructed as
(superpositions of) multi-particle states.
In particular, a ``single particle'' state is given as
\begin{equation}
 | \Bphi \rangle =
\int dx\; \Bpsi^\dagger(x) \Bphi(x) |\mbox{vac} \rangle ,
\end{equation}
where $|\mbox{vac}\rangle$ is the vacuum,
$\Bpsi^\dagger(x) = \Bpsi^T(x)$ for the real field
under consideration here,
and
\begin{equation}
 \Bphi(x) = 
\begin{pmatrix}
  \phi_R(x) \\
  \phi_L(x)
\end{pmatrix}
\label{eq.Bphi_def}
\end{equation}
is a two-component ``wavefunction'' (which is generally complex).
The Hamiltonian can be diagonalized by solving the
eigenequation 
(single-particle Schr\"{o}dinger equation)
\begin{equation}
 \hBdG \Bphi(x)  = \epsilon \Bphi(x) .
\end{equation}
In the infinite line $-\infty < x < \infty$, this
can be solved using the Fourier transformation
\begin{equation}
 \Bphi(x) = \int \frac{dp}{2\pi} \; \Bphi(p) e^{ipx},
\end{equation}
which results in
\begin{equation}
 \hBdG(p) = p \tau^z - m \tau^y .
\end{equation}
This implies that the single-particle eigenstate wavefunctions are
\begin{equation}
 \Bphi^{(+)}(p) \propto
\begin{pmatrix}
 p + \sqrt{p^2+m^2} \\
 - im
\end{pmatrix}
\label{eq.Bphip_pos}
\end{equation}
and
\begin{equation}
 \Bphi^{(-)}(p) \propto
\begin{pmatrix}
 im \\
p + \sqrt{p^2+m^2}
\end{pmatrix}
\end{equation}
which satisfy
\begin{equation}
 \hBdG(p) \Bphi^{(\pm)} (p) = \pm \epsilon(p) \Bphi^{(\pm)} (p),
\end{equation}
where
\begin{equation}
 \epsilon(p) = \sqrt{p^2+m^2}.
\label{eq.dispersion}
\end{equation}
The corresponding single-particle creation operator is
\begin{equation}
 {\eta^{(\pm)}(p)}^\dagger = \Bpsi^\dagger(p) \Bphi^{(\pm)}(p) .
\end{equation}
As a consequence of the real (Majorana) nature of the field
$\Bpsi$, we find
\begin{equation}
 {\eta^{(+)}(p)}^\dagger = \eta^{(-)}(-p) .
\end{equation}
Thus the negative-energy particles are actually identical
to the holes/antiparticles of the positive-energy particles
and not independent degrees of freedom, and
the system can be described solely in terms of positive-energy
particles created by ${\eta^{(+)}(p)}^\dagger$.
(Subtleties concerning possible zero-energy states will be discussed
later.)

In any case, the ``Majorana mass'' $m$ determines the gap
in the single-particle spectrum.
As an effective theory for the Ising model,
$m$ is proportional to the deviation
from the critical point $\Gamma - \Gamma_c$
(see Appendix~\ref{app.Ising_cont}).
The Ising spin ($\sigma^z$) operator is understood as a ``twist field''
with respect to the Majorana fermion field~\cite{CFT-YellowPages}.

In this paper, I will study the universal finite-size scaling of
the energy gap in the nearly-critical
quantum Ising chain with a one-parameter
family of the boundary conditions
\begin{equation}
\calH_I = - J \sum_{j=0}^{N-2}  \sigma^z_j \sigma^z_{j+1}
- b J \sigma^z_{N-1} \sigma^z_0
- \Gamma \sum_{j=0}^{N-1} \sigma^x_j ,
\label{eq.qIsing_b}
\end{equation}
where $N = L/a$ is the number of spins and $b$ is the
parameter representing the strength of the ``defect''.
This includes the periodic ($b=1$), antiperiodic ($b=-1$),
and open ($b=0$) boundary conditions as special cases.
This model corresponds to an anisotropic limit of the
two-dimensional classical Ising model with a defect line,
and they share the same universal critical behavior
near the critical point.
These models have been studied
with various
methods~\cite{Bariev1979,McCoyPerk1980,Kadanoff-Defect_1981,Brown1982,
Turban1985,GuimaraesFelicio1986,HenkelPatkos1987,
CabreraJullien_PRL1986,ZinnJustin_PRLC1986,
CabreraJullien_PRB1987,AbrahamKoSvrakic1989,
HenkelPatkosSchlottmann1989,Grimm1990,
Delfino-Mussardo-Simonetti_1994,OA-IsingDefect-PRL,OA-IsingDefect-NPB}.
In particular, at the critical point, 
this one-parameter family of the boundary conditions
corresponds to a ``fixed line'' with continuously changing boundary
scaling dimensions.
The finite-size energy spectrum in the scaling limit at the critical
point can be described in terms of boundary CFT.
The purpose of this paper is to extend such a field-theory description
of the universal scaling of the finite-size energy spectrum
in the massive (off-critical) regime.

A central issue in the present problem is that,
even though the effective field theory appears symmetric
with respect to the mass inversion $m \to -m$,
the original quantum Ising model is not.
The mass inversion is a mapping between the ordered
($\Gamma < \Gamma_c$) and disordered ($\Gamma > \Gamma_c$) phases,
which are certainly inequivalent.
The ``symmetry'' with respect to the mass inversion
corresponds to the
Kramers-Wannier duality~\cite{KramersWannier,Kogut-RMP1979}.
Even though the duality exists in a certain sense, it does not
mean that the spectrum is identical between the two cases.
In fact, the ground state is
unique in the disordered phase
( $m > 0$ or $\Gamma > \Gamma_c$) and gapped,
while the lowest finite-size gap asymptotically vanishes
in the limit of a large system size in the ordered phase
( $m<0$ or $\Gamma < \Gamma_c$) reflecting the spontaneous
symmetry breaking.
I emphasize that, although the finite-size energy spectrum
depends on the boundary condition,
the asymmetry with respect to the mass inversion
exists for any boundary condition,
as it is the bulk that dictates the spontaneous symmetry breaking.
Understanding of the asymmetry in terms of the relativistic
(Majorana fermion) field theory is an interesting problem.

Not surprisingly, many studies on the finite-size spectrum
of the quantum Ising chain have been reported in the past.
Pfeuty discussed the finite-size energy spectrum with the open
boundary condition and its asymptotic behavior in the
thermodynamic limit~\cite{Pfeuty_1970}.
Abraham studied the spectrum of the transfer matrix
of the two-dimensional classical Ising model on a strip
of a finite width with the open (free) boundary
condition~\cite{Abraham_1971}.
As noted above, although these are microscopically different objects,
the universal critical behavior of these
spectra should be identical.
In Refs.~\cite{Pfeuty_1970,Abraham_1971},
a single-particle state characterized by an imaginary momentum
was found in the ordered phase.
This leads to the asymptotic double degeneracy of the ground states,
which reflect the spontaneous breaking of the Ising spin flip
($\mathbb{Z}_2$) symmetry.

More recently, Kitaev~\cite{Kitaev-chain} discussed the
energy spectrum of the Kitaev chain with the open boundary condition.
For the open boundary condition, the energy spectrum is
identical between the quantum Ising chain and the Kitaev chain.
In the topological phase, Kitaev found
the Majorana zero modes localized near the ends, and
the associated quasi-degeneracy of the ground states.
These are mathematically identical~\cite{Fendley-Parafermion_2012}
to what were found in Refs.~\cite{Pfeuty_1970,Abraham_1971}.
However, the physical significance of the Majorana zero modes
and its potential relevance for topological quantum computation
were not recognized until pointed out by Kitaev~\cite{Kitaev-chain}.

For the periodic boundary condition,
Hamer and Barber~\cite{HamerBarber-Ising1981}
studied the lowest gap (among other things) in the scaling limit
from the field-theory viewpoint.
However, apparently they did not pay much attention to the asymmetry 
with respect to the mass inversion
(asymmetry between the ordered and disordered
phases).
On the other hand, Sachdev~\cite{Sachdev-Ising-crossover}
discussed the spin-spin correlation function in an infinite
Ising chain at finite temperature, and obtained the exact universal
scaling function for the correlation length in the scaling limit.
As one can see from the Quantum Transfer Matrix
formalism~\cite{Koma-ThermalBA,Kluemper-QTM},
the universal scaling of the inverse correlation length for
the spin-spin correlation function at finite temperature should be
identical to that of the lowest gap in a finite size chain,
if the system is Lorentz invariant in the low-energy limit.
Thus Sachdev's formula can be regarded as the universal scaling
function of the lowest gap.
Indeed, it clearly exhibits the asymmetry with respect
to the mass inversion. 

The finite-size energy spectrum of the quantum Ising chain
(or transfer matrix spectrum of the classical Ising model on a strip)
with the one-parameter family of the boundary conditions~\eqref{eq.qIsing_b}
has been also discussed
in Refs.~\cite{
CabreraJullien_PRL1986,ZinnJustin_PRLC1986,CabreraJullien_PRB1987,
BarberCates_1987,PrivmanSvrakic_PRL1989,PrivmanSvrakic_JSP1989,
AbrahamKoSvrakic1989}.
In particular, a thorough analysis of the spectrum in the ordered phase
based on the exact solution
was presented by Abraham, Ko, and \v{S}vraki\'{c}~\cite{AbrahamKoSvrakic1989}.
However, the universal scaling function of the gap,
including the universal crossover between the ordered and disordered phase,
has not been discussed in the full detail.

In this paper, I will discuss the universal scaling function
of the lowest gap,
for the general one-parameter family of boundary conditions.
The field-theory analysis in this paper exactly reproduces the result of
Ref.~\cite{Sachdev-Ising-crossover} for the special case
of the periodic boundary condition, although the approach is
rather different.
I emphasize that, while the exact solution of the quantum Ising chain
depends on the particular lattice model~\eqref{eq.qIsing_b},
the universal scaling function
should apply to any nearly critical system which belongs to the same
Ising universality class.

The quantum Ising chain is mapped to the Kitaev chain with some
constraints on the parameters.
The constraints also apply to the defect.
Thus, the Kitaev chain allows somewhat more general defects than
the quantum Ising chain, as discussed in
Refs.~\cite{Navaetal-Majorana2017,KKWK2017}.
As a byproduct of the present study, I analyse the generalized
defect in the Kitaev chain in the vicinity of a quantum critical
point, based on the relativistic Majorana fermion field theory. 
I give a field-theory picture for the
most remarkable observation in Ref.~\cite{KKWK2017} that
Majorana zero modes exist near the defect under a certain
condition, even when the defect coupling is non-vanishing.

This paper is organized as follows.
In Sec.~\ref{sec.open}, I will discuss the universal scaling function
of the lowest gap in finite chains with the open boundary condition,
for which bound states localized around the chain ends appear
in the ordered phase.
The universal scaling function for the periodic boundary condition
is discussed in Sec.~\ref{sec.PBC}. An effective reversal of
the fermion number parity leads to the asymmetry between of
the lowest gap in the ordered and disordered phases.
In Sec.~\ref{sec.general}, the universal scaling function of
the lowest gap is discussed for the more general
one-parameter family of
the boundary condition. Both the bound states and the fermion
number parity reversal play roles. 
Finally, Sec.~\ref{sec.conclusion} is devoted to conclusions
and discussions.

Derivation of the relativistic
Majorana fermion field theory as the effective
theory of the quantum Ising chain is presented
in Appendix~\ref{app.Ising_cont}.
In Appendix~\ref{app.scatt_defect}, scattering of Majorana fermions
at the defect is analyzed, in order to determine
the parameters in the field theory description in terms of
the lattice model.
Scattering at  more general class of defects in the Kitaev chain is discussed
in Appendix~\ref{app.Kitaev_scatt}.

\section{Open boundary condition}
\label{sec.open}

Here I consider the quantum Ising chain with the open boundary condition.
In this case, the Jordan-Wigner transformation is well defined.
Thus, the quantum Ising chain is mapped to the Kitaev chain with the
open boundary condition, and the spectrum of the two models are completely
identical.
In the vicinity of the quantum critical point,
the effective Majorana fermion field theory is applicable.
The open boundary condition for the Kitaev chain is translated to
the boundary conditions
\begin{align}
 \psi_R (x=0) &= \psi_L (x=0) ,
\label{eq.psi-bc-0}
\\
 \psi_R (x=L) &= - \psi_L (x=L) ,
\label{eq.psi-bc-L}
\end{align}
for the Majorana fermion fields in the chiral basis.
The boundary conditions on the fields can be simply translated
to those on the wavefunctions~\eqref{eq.Bphi_def}:
\begin{align}
 \phi_R (x=0) &= \phi_L (x=0) ,
\label{eq.phi-bc-0}
\\
 \phi_R (x=L) &= - \phi_L (x=L) .
\label{eq.phi-bc-L}
\end{align}
These boundary conditions were obtained in Ref.~\cite{CSRL-SPT-BCFT}
for massless case, but this is also valid for non-vanishing mass.
This is natural, since within length-scales shorter than the inverse mass,
the massive system should behave similarly to the massless CFT.
The relative sign difference between the boundary conditions at two ends
was related to the conformal spin of the fermion field~\cite{CSRL-SPT-BCFT}
in terms of CFT.
Alternatively, these boundary conditions can be derived 
from an explicit calculation.
Setting $b=0$ ($\chi=\pi/2$) in the transmission/reflection amplitudes
Eqs.~\eqref{eq.Tcoeff},~\eqref{eq.RRcoeff}, and~\eqref{eq.RLcoeff},
which are obtained later in Sec.~\ref{sec.general}
and Appendix~\ref{app.scatt_defect} for the generalized defect boundary
conditions,
I find $T_{\rightarrow}=T_{\leftarrow}=0$, $R_{\leftarrow}= 1$,
and $R_{\rightarrow}=-1$ for the open boundary conditions.
The last two (reflection) amplitudes correspond to
Eqs.~\eqref{eq.phi-bc-0} and~\eqref{eq.phi-bc-L}.

The Schr\"{o}dinger equation for the Majorana fermion field
in the real space
\begin{align}
 \hBdG \Bphi = 
\begin{pmatrix}
 - i \partial_x & i m \\
 -i m & i \partial_x 
\end{pmatrix}
\begin{pmatrix}
 \phi_R \\
 \phi_L
\end{pmatrix}
= \epsilon
\begin{pmatrix}
 \phi_R \\
 \phi_L
\end{pmatrix}
\label{eq.Maj-Schrodinger}
\end{align}
together with the boundary condition~\eqref{eq.phi-bc-0}
admits a potential zero-energy bound-state solution
\begin{align}
\Bphi(x)
\propto
\begin{pmatrix}
 1  \\
 1
\end{pmatrix}
e^{m x},
\end{align} 
near the left end of the chain $x \sim 0$.
For this solution to be normalizable (and be a bound-state),
\begin{equation}
 m < 0,
\label{eq.negative-m}
\end{equation}
has to be satisfied.
Likewise, Eq.~\eqref{eq.Maj-Schrodinger} together with the
the boundary condition~\eqref{eq.phi-bc-L} admits
a zero-energy bound state
\begin{align}
\Bphi(x)
\propto
\begin{pmatrix}
 1  \\
 - 1
\end{pmatrix}
e^{m (L -x)},
\end{align} 
localized near the right end of the
chain $x \sim L$, under the same condition~\eqref{eq.negative-m}.
(Here we pretend $L$ to be ``infinite'' and treat the two
bound states separately. For a finite $L$, in reality, the
``bound state'' has to satisfy the boundary conditions on the
both ends and its energy is lifted from zero, as I will discuss
later.)

Thus, the Kitaev chain has two ``Majorana zero modes'' localized
near the both ends of the chain in the
``topological phase''~\eqref{eq.negative-m}.
As it was pointed out in the celebrated paper
by Kitaev~\cite{Kitaev-chain},
the two Majorana zero modes constitute a single ``qubit'',
so that they can be regarded as a spatially distributed qubit
which is robust against decoherence.
The presence of the Majorana zero modes was demonstrated~\cite{Kitaev-chain}
in the exactly solvable
limit of $t=|\Delta|$ and $\mu=0$,
where the zero modes are completely localized.
In this limit, the ground states are exactly two-fold degenerate for
any length of the chain.
At generic points in the topological phase,
the two ground states are only nearly degenerate, split by the
exponentially small energy $\sim e^{-mL}$.
As I have mentioned earlier, the energy spectrum of the quantum Ising chain
is identical to that of the Kitaev chain under the open boundary
condition.
Thus, there are two quasi-degenerate ground states only in one side
of the critical point, Eq.~\eqref{eq.negative-m}.
This is nothing but the ordered phase of the quantum Ising chain,
and the quasi-degeneracy of the ground states signals the spontaneous
breaking of the \Ztwo symmetry.
As noted in the Introduction, the ``Majorana zero modes'' 
had been mathematically discovered in the Ising model
context earlier~\cite{Pfeuty_1970,Abraham_1971}.

Now let me discuss the universal scaling function of the smallest gap
(splitting of the quasi-degenerate ground states in the topological/ordered
phase) based on the effective Majorana fermion field theory.
The two independent ``plane wave'' eigenstates with the energy
$+ \sqrt{p^2 + m^2}$ in the real space are
\begin{align}
 \Bphi^{(+)}_p(x) & \equiv \Bphi^{(+)}(p) e^{i p x} ,
\\
 \Bphi^{(+)}_{-p}(x) & \equiv
\begin{pmatrix}
 i m  \\
 p + \sqrt{p^2 + m^2}
\end{pmatrix}
e^{i p x}
\notag \\
& \propto \Bphi^{(+)}(-p) e^{-i p x},
\end{align}
where $\Bphi^{(+)}(p)$ is defined in Eq.~\eqref{eq.Bphip_pos},
for $p \geq 0$.
I note that, for $p=0$, these two states reduce to the single state
\begin{align}
 \Psi^{(+)}(p=0) &= 
\begin{pmatrix}
 1  \\
-i \; \sgn{m}
\end{pmatrix}
.
\end{align}
However, since this does not satisfy the open boundary 
conditions, $p \neq 0$ can be assumed
(except for the special case when $p=0$ becomes a double root).

For $p \neq 0$, the single-particle energy eigenstate  can be
written as the superposition
\begin{equation}
\Bphi^{(+)}(x) = a_+  \Bphi^{(+)}_p(x) + a_-  \Bphi^{(+)}_{-p}(x)
\label{eq.Psi_superpos}
\end{equation}
The boundary condition at $x=0$ then requires
\begin{equation}
 (1, -1)^T 
\Bphi^{(+)}(x=0) =
\left( p + im + \sqrt{p^2+m^2} \right)
a_+ 
-
\left( p - im + \sqrt{p^2+m^2} \right)
a_-
= 0 . 
\end{equation}
This implies
\begin{align}
 \frac{a_+}{a_-} = \frac{p-im + \sqrt{p^2+m^2}}{p+im+\sqrt{p^2+m^2}}
 = \frac{p - i m}{\sqrt{p^2+m^2}} .
\end{align}
Similarly, the boundary condition at $x=L$ requires
\begin{align}
 \frac{a_+}{a_-} = - \frac{p + i m}{\sqrt{p^2 + m^2}} e^{-2 i p L} .
\end{align}
The two boundary conditions thus imply
\begin{equation}
 \frac{ (p- im)^2}{p^2+m^2} e^{2 i p L} = -1 .
\end{equation}
In the massless limit $m=0$ which corresponds to the critical point,
this is equivalent to $e^{2 i p L}=-1$, which implies
\begin{align}
 p &= \frac{2n - 1}{2 L} \pi ,
\label{eq.p_for_m0}
\end{align}
where $n$ is a positive integer.
This recovers the result of Ref.~\cite{CSRL-SPT-BCFT} that the fermions
obey the antiperiodic boundary condition on a ring of the
doubled length $2L$.

Also for $m \neq 0$, there are infinite number of solutions.
The condition to be satisfied by the momentum $p$ can be rewritten as
\begin{align}
 \tan{pL} = - \frac{p}{m} ,
\label{eq.cond_tanpL}
\end{align}
excluding the trivial solution $p =0$ (except for the
special case of the double root).
The solutions~\eqref{eq.p_for_m0} for $m=0$ corresponds
to the poles of $\tan{pL}$.
For a finite $m$, the solutions will be shifted from the poles of $\tan{pL}$.
Graphically, they can be identified as crossing points between the
curve $\tan{pL}$ and the straight line $- p/m$.
For $m <0$, the crossing point at the smallest $p>0$ decreases
as $|m|$ is increased.
Eventually, at
\begin{align}
 - m L = 1,
\label{eq.mL_eq_1}
\end{align}
the smallest root $p$ merges to $0$.
While the root $p=0$ is usually excluded, it should be taken into
account in this ``double root'' case.

For 
\begin{align}
 - m L > 1,
\label{eq.mL_gt_1}
\end{align}
the real solution corresponding to the smallest $p$ disappears.
However, it can be regarded that the solution turned pure imaginary
in this case. That is,
\begin{align}
 p = i \lambda,
\label{eq.p_imag}
\end{align}
where $\lambda$ is a real root satisfying
\begin{align}
 \tanh{\lambda L} = - \frac{\lambda}{m} .
\end{align}
This equation indeed has a unique real solution (excluding $\lambda=0$)
when~\eqref{eq.mL_gt_1}.

The imaginary solution reflects the presence of the bound states
localized near the end.
As discussed in Sec.~\ref{sec.Introduction}, the bound states are
at zero energy in the limit of infinitely long chain.
However, the energy is lifted to a positive value in a finite chain,
reflecting the mixing between the two bound states.

The effective Hamiltonian of the system takes the form of
Eq.~\eqref{eq.qIsingHcont2}, where the parameter $p$
runs over the solution of Eq.~\eqref{eq.cond_tanpL}
excluding $p=0$ and including the possible pure imaginary solution.
The ground state of the system is nothing but the ``vacuum'' in terms
of the $\eta_p$-fermions.
The lowest excited state is given by creating a fermion with the
smallest possible $p$, which we denote $p^*$.
When Eq.~\eqref{eq.mL_gt_1} holds, $p^*$ is understood as the
pure imaginary solution.

The lowest excitation gap $\epsmin$ in the finite-size chain with the open
boundary condition is then given as
\begin{equation}
 \epsmin = \sqrt{{p^*}^2+m^2} .
\label{eq.epsmin_open}
\end{equation}
Introducing the scaling variable
\begin{equation}
s \equiv mL,
\label{eq.def_s}
\end{equation}
and the scaled lowest gap
\begin{equation}
 \Delta \equiv L \epsmin ,
\label{eq.def_Delta}
\end{equation}
the result~\eqref{eq.epsmin_open} reads
\begin{equation}
 \Delta = \sqrt{{q^*}^2 + s^2}
\end{equation}
where $q^*$ is the smallest positive solution of
\begin{equation}
 \tan{q} = - \frac{q}{s} ,
\end{equation}
or its pure imaginary solution when $s < -1$.

When $s<-1$,
the scaled lowest gap may be written as
$\Delta = - s \sqrt{\rho}$, where $\rho$ is
the positive solution of
\begin{align}
 \tanh{\left( |s| \sqrt{1 - \rho} \right)} =  \sqrt{1 - \rho} .
\label{eq.scaledgap_obc}
\end{align}
Deep inside the ordered phase, $s <0$ and $|s|$ is large.
This requires $\rho \sim 0$
(namely $\epsmin \sim 0$), indicating the
quasi-degeneracy of the ground states.
More precisely, $\rho$ can be expanded in terms of the
small parameter $\zeta = e^{-2 |s|}$.
As I will show below, $\rho$ is $O(\zeta)$.
Here I obtain $\rho$ up to $O(\zeta^2)$.
Up to this order, $\rho^3 = O(\zeta^3)$ and hence can be ignored.
Expanding Eq.~\eqref{eq.scaledgap_obc} up to the second order in $\rho$, 
\begin{equation}
 \left(2|s|^2 \tanh^3{|s|} + l \tanh^2{|s|} - 2|s|^2 \tanh{|s|} -l +1 \right)
\rho^2
+ 4 (|s| \tanh{|s|}^2 -|s| +1 ) \rho
+ \left( \tanh{|s|}-1 \right) = 0.
\end{equation}
The solution satisfying $\rho>0$ for small $\zeta$ reads,
up to $O(\zeta^2)$,
\begin{align}
 \rho = 4 \zeta + (16 l - 8) \zeta^2 + O(\zeta^3) .
\end{align}
It follows that
\begin{equation}
 \Delta = 2 |s| \left[ e^{- |s| } + (2 |s| - 1) e^{-3 |s|} \right]
+ O(e^{-5 |s|}),
\end{equation}
or, in terms of the original variables,
\begin{align}
 \epsmin = 2 |m| \left[
e^{- |m| L} + (2 |m| L - 1) e^{-3 |m| L} \right] + O(e^{-5 |m| L}),
\
\end{align}
in the ordered region $m<0$ and $|m|L \gg 1$.
While the first term in the expansion is well known, the second term
is also universal.
It would be also possible to obtain higher-order terms in this
expansion of the finite-size gap (splitting of the quasi-degenerate
ground states).

\section{Periodic boundary condition}
\label{sec.PBC}

Here I discuss the universal scaling of the finite-size gap
of the Ising chain under the periodic boundary condition.
In this case, there are no Majorana bound states since there is no
open boundary to localize such states.
Nevertheless, physically one should expect the quasi-degenerate
ground states in the ordered phase $m<0$.
In fact there is a different mechanism to ensure this
for the periodic boundary condition, which I shall demonstrate
in the following.

The Jordan-Wigner transformation maps the quantum Ising chain
to fermionic (Kitaev) chain only with local (short-ranged) hoppings
and interactions.
However, under the periodic boundary condition,
the Jordan-Wigner transformation of 
the Ising interaction at the ``boundary'' leads to a non-local
term containing the Jordan-Wigner string wrapping around
(almost) the entire system.
Fortunately, such a term can be written in terms of the total
fermion number parity and a local Hamiltonian.
Since the fermion number parity is conserved, it can be regarded
as a constant.
This leads to an interesting situation: the resulting fermion model
is still ``free'', but the boundary condition for the fermion
depends on the total fermion number parity.
When $(-1)^F = 1$ (even fermion number) fermions obey the antiperiodic
boundary condition, while they obey the periodic boundary condition
when $(-1)^F= -1$ (odd fermion number).

It turns out that the ground state belongs to the even fermion number
sector, while the first excited state does to the odd fermion number
sector.
The ground state is given by the ``vacuum'' in terms of fermions.
Thus the ground-state energy reads
\begin{equation}
E_0 =  - \frac{1}{2} \sum_{p = \frac{\pi (2n -1)}{L}} \epsilon(p) ,
\end{equation}
where I included the ``zero-point'' energy of each mode
(which is essential as it can be seen below).
Here $p$ runs over the quantized momenta $\pi (2n-1)/L$ with
$n \in \mathbb{Z}$, corresponding to the antiperiodic boundary condition.

On the other hand, the first excited state belongs to the odd
fermion number sector.
In order to be in the odd sector, one has to create odd number of
fermions on the vacuum (however, see below).
It is obvious that the first excited state, the lowest energy among
such states, is given by creating just one fermion with the minimum
energy at $p=0$.
Since the energy of the fermion at $p=0$ is given as $\sqrt{m^2} = |m|$,
the energy of the first excited state appears to be given as
\begin{equation}
E_1 =  |m| - \frac{1}{2} \sum_{p = \frac{2 \pi n}{L}} \epsilon(p) ,
\label{eq.E1PBC0}
\end{equation}
where I included the zero-point energy for the periodic boundary
condition.
However, this expression contradicts the physics of the Ising chain,
since the energies $E_0$ and $E_1$ are now completely symmetric
with respect to $m \to - m$ and do not show the expected
difference between the ordered and disordered phases.
In particular, in the thermodynamic limit $L \to \infty$ of
an off-critical chain ($m \neq 0$), the zero-point energies are
asymptotically the same and the finite-size gap reads
\begin{equation}
 \Delta = E_1 - E_0 \sim |m| .
\end{equation}
Namely, the expected quasi-degeneracy of the ground states is
absent in the ordered phase $m<0$.

The resolution to this puzzle is given as follows.
Let us start from the disordered phase $m>0$, where
the expression~\eqref{eq.E1PBC0} is indeed valid.
As we decrease the mass parameter $m$, the energy $m$ of the
lowest mode $p=0$ decreases.
When we cross the quantum critical point $m=0$ and enter
the ordered phase $m<0$, from the dispersion relation
it appears that the energy of the lowest mode is given as $|m|$.
However, in a finite-size system, there should be no singularity
and thus the energy of the lowest mode cannot exhibit the
cusp-like dependence as $|m|$ on the parameter $m$ at $m=0$.
In this respect, the energy of the lowest mode $p=0$ should
be understood as $m$, not $|m|$.
This energy becomes negative in the ordered phase $m<0$.
For convenience, we can always apply the particle-hole transformation
to such a negative-energy mode so that the single-particle
energy is non-negative.
This particle-hole transformation applied only to $p=0$ recovers
the standard dispersion relation~\eqref{eq.dispersion}.
Usually this does not leave any other side-effects, but here
we have to be careful with the fermion number parity constraint.
Since the ``occupied'' and ``unoccupied'' states are exchanged
for $p=0$, the original fermion number parity $(-1)^F$
and the fermion number parity $(-1)^{\tilde{F}}$ (in the new
basis for which Eq.~\eqref{eq.dispersion} holds) are different
and are related by
\begin{equation}
 (-1)^F = - (-1)^{\tilde{F}} .
\end{equation}
In particular, the first excited state of the system,
that is the lowest-energy state in the ``odd sector'',
is given by one fermion with the lowest energy $m$ at $p=0$
in the original representation.
Applying the particle-hole transformation, the dispersion
relation recovers the form~\eqref{eq.dispersion} while
the first excited state now becomes the vacuum for
fermions obeying the periodic boundary condition.
(Of course, this conclusion can be obtained by
performing the Jordan-Wigner transformation carefully,
without resorting to the analyticity argument.)

\begin{figure}
\begin{center}
\includegraphics[width=0.6\textwidth]{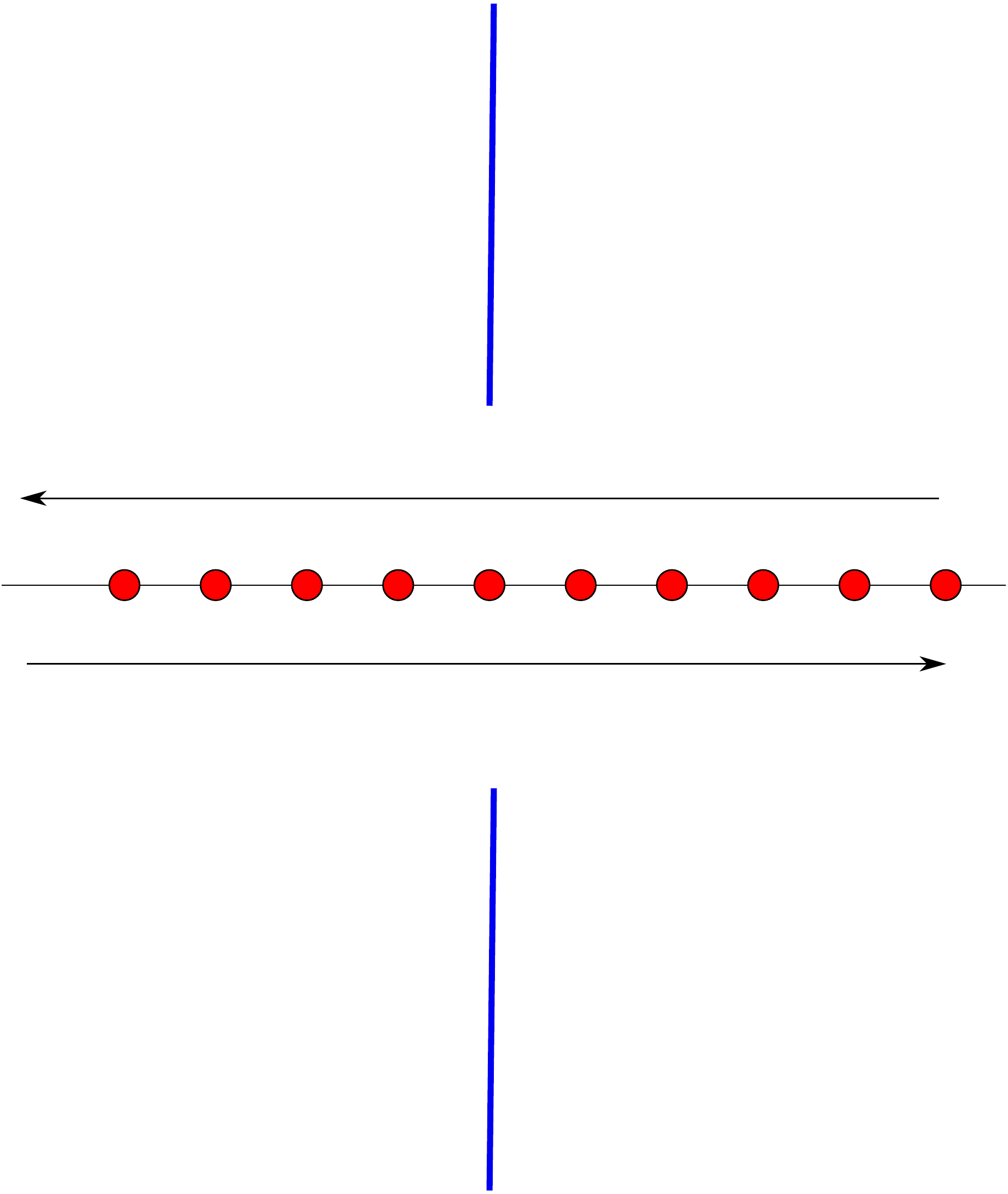}
\caption{The contour $C$. The blue lines represent branch cuts.}
\label{fig.contour1}
\end{center}
\end{figure}

\begin{figure}
\begin{center}
\includegraphics[width=0.6\textwidth]{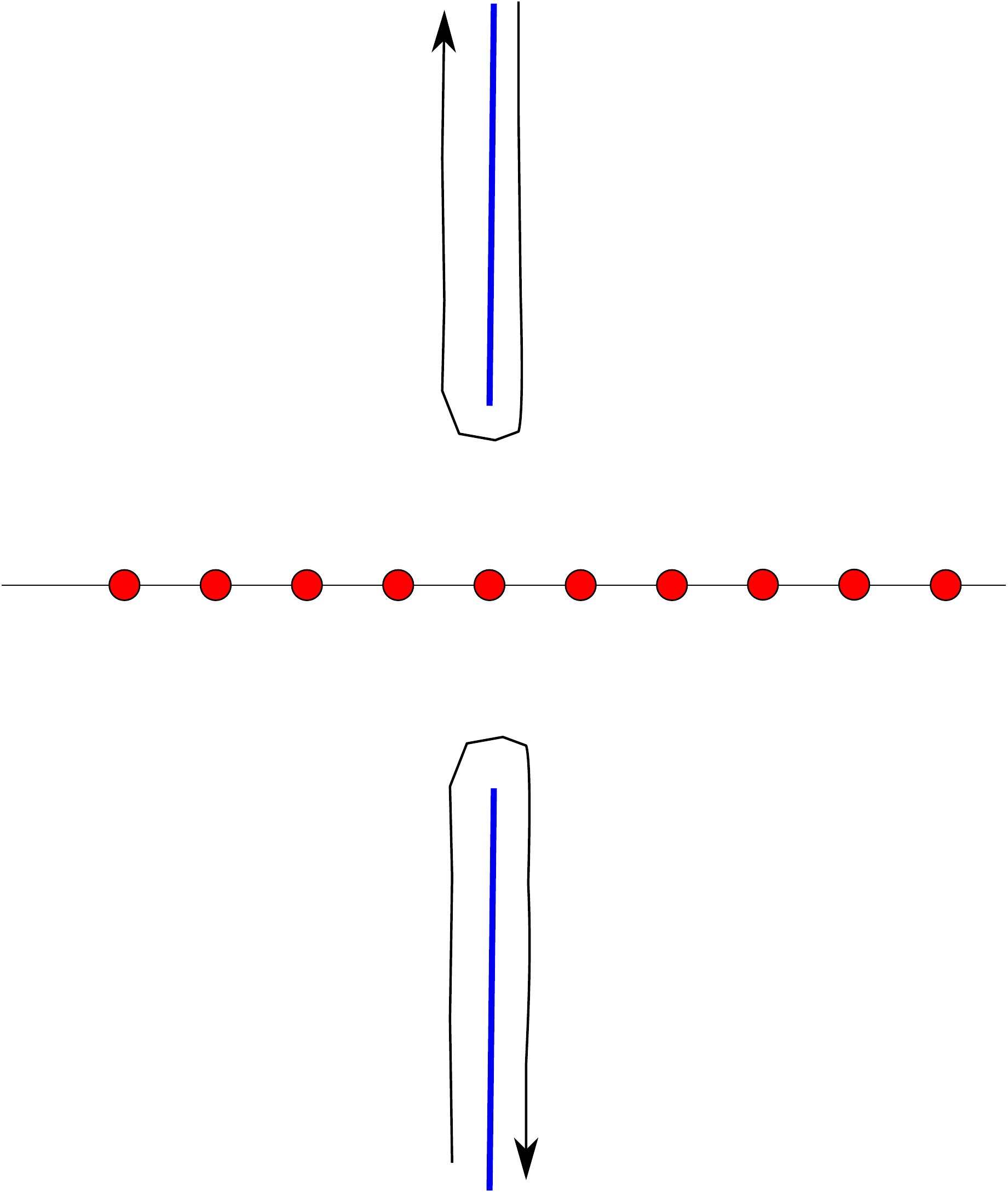}
\caption{The deformed contour}
\label{fig.contour2}
\end{center}
\end{figure}

This analysis leads to the correct expression of the energy
of the first excited state
\begin{equation}
E_1 =  m \Theta{(m)}
   - \frac{1}{2} \sum_{p = \frac{2 \pi n}{L}} \epsilon(p) ,
\label{eq.E1PBC}
\end{equation}
where $\Theta$ is the Heaviside step function
\begin{equation}
 \Theta(m) \equiv
\begin{cases}
 1 & (m>0), \\
 0 & (m \leq 0) .
\end{cases}
\end{equation}
The first gap thus reads
\begin{equation}
\epsmin =  m \Theta(m) +
\frac{1}{2}
\left[
 \sum_{p=\frac{2\pi n}{L}}  - \sum_{p=\frac{\pi(2n-1)}{L}} 
 \sqrt{p^2+m^2}
\right] .
\label{eq.gapPBC}
\end{equation}
While the zero-point energy in each of $E_0$ and $E_1$ is divergent,
their difference is finite, and
can be written as a contour integral
\begin{equation}
I =
\frac{L}{4 \pi i} \oint_C \frac{\sqrt{m^2+ p^2}}{\sin{p L}} \; \mathrm{d}p ,
\end{equation}
where the contour $C$ is defined as in Fig.~\ref{fig.contour1}.
The branch cut due to the square root should run from
$p = \pm im$ to infinity, so that it does not intersect with
the contour $C$.
We can simply take the branch cut parallel to the imaginary axis.
Deforming the contour so that it surrounds the branch cut
as in Fig.~\ref{fig.contour2},
we find
\begin{equation}
I = \frac{L}{\pi} \int^{\infty}_{|m|}
\frac{\sqrt{\kappa^2 - m^2}}{\sinh{\kappa L}} \; \mathrm{d}\tilde{\kappa} .
\end{equation}

Changing the integration variable to $y=\kappa L$,
the scaled lowest gap~\eqref{eq.def_Delta}
can be written as a universal function of
the scaling variable~\eqref{eq.def_s} as
\begin{equation}
\Delta = 
\left[ s \Theta(s) +
\frac{1}{\pi}\int^{\infty}_{|s|} \frac{\sqrt{y^2-s^2}}{\sinh{y}} \; dy
\right].
\label{eq.gapPBCfinal}
\end{equation}
This universal gap function was actually obtained more than
20 years ago by Sachdev~\cite{Sachdev-Ising-crossover}.
He studied the universal scaling function for the spin-spin
correlation function in the quantum Ising chain at finite temperature.
Since the spin-spin correlation function is the most long-ranged,
its inverse correlation length corresponds to the lowest gap
in the spectrum of the quantum transfer matrix.
As the scaling (continuum) limit of the quantum Ising chain is
described by the Lorentz-invariant field theory, the scaling limit
of the quantum transfer matrix is identical to the Hamiltonian.
Thus the universal scaling function for the spin-spin correlation
length at finite temperatures is identical to the universal
gap scaling function I have derived here.
Indeed, if the system size $L$ is replaced by the inverse temperature,
Eq.~\eqref{eq.gapPBCfinal} agrees exactly with
the expression in Ref.~\cite{Sachdev-Ising-crossover}, which
was derived by a rather different method (taking the scaling limit
of the Toeplitz determinant expression).
The origin of the non-analytic term $s \Theta(s)$ 
is perhaps more naturally understood in the present formalism.
Despite the non-analytic nature of the step function,
the entire scaling function must be analytic in $s$,
as pointed out in Ref.~\cite{Sachdev-Ising-crossover}.
Namely, a non-analytic contribution from the integral
around $s=0$ cancels the non-analyticity originating from
the step function.

\section{General boundary conditions}
\label{sec.general}

Now I consider the spectrum of the quantum Ising chain 
with the defect as in Eq.~\eqref{eq.qIsing_b}.
While some of the results in this section could be
found by taking the scaling limit of those
in Ref.~\cite{AbrahamKoSvrakic1989}, I will elucidate
interesting structures appearing in the crossover
between the ordered and disordered sides.

At the bulk critical point,
a continuously changing ``boundary'' critical
exponent~\cite{Bariev1979,McCoyPerk1980,Kadanoff-Defect_1981,Brown1982,
Turban1985,GuimaraesFelicio1986,HenkelPatkos1987,
CabreraJullien_PRL1986,ZinnJustin_PRLC1986,
CabreraJullien_PRB1987}
as a function of $b$
is found for the defect model~\eqref{eq.qIsing_b}.
In terms of CFT,
this can be understood, after a folding procedure,
as a dependence on the boundary value $\varphi_0$
of the $\mathbb{Z}_2$ orbifold of the
free boson field theory~\cite{OA-IsingDefect-PRL,OA-IsingDefect-NPB}.
However, in order to extend the analysis to the massive case,
it is more convenient
to treat the defect in terms of the fermions.
In the continuum (scaling) limit, the defect is represented
as a localized mass term for the relativistic Majorana fermions.
Such a defect causes elastic
scatterings (transmission and reflection) of the relativistic
Majorana fermions, and can be characterized
by the transmission/reflection coefficients $T$ and
$R$~\cite{Delfino-Mussardo-Simonetti_1994}.
They are functions of the magnitude of the momentum of the incoming
fermion, and given as
\begin{align}
 T_{\rightarrow} &= T_{\leftarrow} =
    \frac{p \cos{\chi}}{p+ i m \sin{\chi}}
\label{eq.Tcoeff}
\\
 R_{\rightarrow} &
    = - \frac{p+im}{p+ i m \sin{\chi}} \sin{\chi},
\label{eq.RRcoeff}
\\
 R_{\leftarrow} &
    =  \frac{p-im}{p+ i m \sin{\chi}} \sin{\chi},
\label{eq.RLcoeff}
\end{align}
where the subscript $\rightarrow$ and $\leftarrow$
represent the direction of the incoming fermion,
and $p>0$ is the absolute value of its momentum.
$\chi$ is the phase shift parameter specifying the universal
scattering property of the defect.
It is determined by the defect strength in the microscopic model.

It should be noted that, as in the case of the periodic boundary
condition, boundary condition with respect to the fermion
depends on the total fermion number parity.
Thus the phase shift parameter generally differs between the even fermion
number sector and the odd fermion number sector, for
a given defect strength $b$.
In the analysis in Appendix~\ref{app.scatt_defect},
I find
\begin{equation}
 \chi = \frac{\pi}{2} + 2 \tan^{-1}{b}.
\label{eq.chi_in_b}
\end{equation}
in the even fermion sector.
By comparing the spectra, the parameter $\chi$ is related
to the boundary value $\varphi_0$ in the orbifold
CFT formulation~\cite{OA-IsingDefect-PRL,OA-IsingDefect-NPB} as
\begin{equation}
\chi = \frac{3\pi}{2} - 2 \varphi_0 .
\end{equation}
In the odd fermion number sector, $\chi$ is replaced by
\begin{equation}
\tilde{\chi} \equiv \pi - \chi, 
\label{eq.def_tildechi}
\end{equation}
which maps $T \to -T$ and $R \to R$.
Values of $\chi$ and $\tilde{\chi}$
for some representative cases are listed
in Table~\ref{tab.chi_b}.

\begin{table}
\begin{tabular}{|c|c|c|c|c|}
$b$
&
$\chi, \tilde{\chi}$
&
bound state
&
$\epsilon^*$ reversal
\\
\hline
\hline
\multirow{2}{*}{
\begin{tabular}{c}
$b=+\infty$ 
\\
infinite ferromagnetic
\end{tabular}
}
&
even $\chi = \frac{3\pi}{2}$ & $m L > 1$ & none 
\\
& odd  $\tilde{\chi} = - \frac{\pi}{2}$ & $m L > 1$ & always
\\ \hline
\multirow{2}{*}{
\begin{tabular}{c}
$1 < b < +\infty$ 
\\
strong ferromagnetic
\end{tabular}
}
& 
even $\pi < \chi < \frac{3\pi}{2}$ &
$ mL > - \tan{\frac{\chi}{2}} > 0$ & none
\\
& odd $ 0 > \tilde{\chi} > - \frac{\pi}{2} $ &
$m L > - \tan{\frac{\tilde{\chi}}{2}} > 0 $ &
$ m L < \log{\frac{\cos{\tilde{\chi}}}{1+\sin{\tilde{\chi}}}}$ 
\\ \hline
\multirow{2}{*}{
\begin{tabular}{c}
$b=1$
\\
periodic
\end{tabular}
}
& 
even $\chi = \pi$ & none & none \\
& odd $\tilde{\chi}=0$ & none & $m<0$ ($p=0$)
\\ \hline
\multirow{2}{*}{
\begin{tabular}{c}
$0 < b < 1 $
\\
weak ferromagnetic
\end{tabular}
}
& 
even $\frac{\pi}{2} < \chi < \pi$ &
$ m L < - \tan{\frac{\chi}{2}} < 0$ & none
\\ 
& odd $\frac{\pi}{2} > \tilde{\chi} > 0$
 & $ m L < - \cot{\frac{\chi}{2}} <0$ &
$ m L < \log{\frac{\cos{\tilde{\chi}}}{1+\sin{\tilde{\chi}}}} $ 
\\ \hline
\multirow{2}{*}{
\begin{tabular}{c}
$b=0$
\\
open
\end{tabular}
}
&
even $\chi = \frac{\pi}{2}$ &
$mL < -1$ &none 
\\ 
& odd $\tilde{\chi}= \frac{\pi}{2}$ & $mL < -1$ & none
\\ \hline
\multirow{2}{*}{
\begin{tabular}{c}
$-1 < b < 0$
\\
weak antiferromagnetic
\end{tabular}
}
&
even $ 0 < \chi < \frac{\pi}{2}$ & $m L < - \tan{\frac{\chi}{2}} < 0 $ &
$ m L < \log{\frac{\cos{\chi}}{1+\sin{\chi}}}$ 
\\
& odd $ \pi > \tilde{\chi} > \frac{\pi}{2}$ & $ m L < - \cot{\frac{\chi}{2}}$ &
none 
\\ \hline
\multirow{2}{*}{
\begin{tabular}{c}
$b=-1$
\\
antiperiodic
\end{tabular}
}
&
even $\chi=0$ & none & $m<0$ ($p=0$) \\
& odd $\tilde{\chi}= \pi$ & none &  none \\ \hline
\multirow{2}{*}{
\begin{tabular}{c}
$- \infty < b < -1 $
\\
strong antiferromagnetic
\end{tabular}
}
&
even $- \frac{\pi}{2} < \chi < 0 $ & $ m L > - \tan{\frac{\chi}{2}} >0 $
& $ m L < \log{\frac{\cos{\chi}}{1+\sin{\chi}}} $
\\
& odd $ \frac{3\pi}{2} > \tilde{\chi} > \pi$ & 
$ mL > - \cot{\frac{\chi}{2}} > 1$ & none 
\\ \hline
\multirow{2}{*}{
\begin{tabular}{c}
$b= -\infty$
\\
infinite antiferromagnetic
\end{tabular}
}
&
even $\chi = - \frac{\pi}{2}$ & $m L > 1$ & always
\\
& odd $ \tilde{\chi} = \frac{3\pi}{2}$ & $m L > 1$ & none
\\ \hline
\end{tabular}
\caption{Values of $\chi$, in the even fermion number sector,
for some representative cases. For the odd fermion number sector,
$\chi$ is replaced by $\tilde{\chi}=\pi - \chi$.
}
\label{tab.chi_b}
\end{table}

An energy eigenstate in the presence of the defect may be
written generally as in Eq.~\eqref{eq.Psi_superpos}.
The coefficients have to satisfy
\begin{align}
 a_+ & = T_{\rightarrow} a_+ e^{i p L} + R_{\leftarrow} a_- , \\
 a_- e^{- i p L}& = 
  T_{\leftarrow} a_-  + R_{\rightarrow} a_+ e^{i p L} .
\end{align}
Existence of a non-trivial eigenstate requires
\begin{equation}
 f(p, \chi) = 0,
\label{eq.fpchi0}
\end{equation}
where
\begin{align}
f(p,\chi) & \equiv (p+ im \sin{\chi}) e^{-i p L}
\det{
\begin{pmatrix}
 1 - T_{\rightarrow} e^{i pL} & - R_{\leftarrow} \\
 - R_{\rightarrow} e^{2 i pL} & 1 - T_{\leftarrow} e^{ i pL}
\end{pmatrix}
}
\notag \\
& =
( p - im \sin{\chi} ) e^{i pL} - 2 p \cos{\chi}
+ (p + i m \sin{\chi}) e^{-i pL} ,
\\
&=
2 \left(
p \cos{pL} -  p \cos{\chi} +  m \sin{\chi} \sin{pL}
\right)
\label{eq.fp_def}
\end{align}
where the overall factor was chosen for convenience.
The infinite number of solutions $p$ for Eq.~\eqref{eq.fpchi0}
determine the eigenstates in the presence of the defect.

As a special case, exactly at the critical point, $m=0$,
Eq.~\eqref{eq.fpchi0} reduces to
\begin{equation}
 e^{2 i p L} - 2 \cos{\chi} e^{i p L} + 1 = 0,
\end{equation}
which means
\begin{equation}
 e^{i p L } =  \cos{\chi} \pm \sqrt{\cos^2{\chi} - 1 } = e^{\pm i \chi} .
\end{equation}
This corresponds to a simple shift of the quantized momenta
\begin{equation}
 p = \frac{2\pi n}{L} \pm \chi ,
\end{equation}
where $p >0$ by definition. 
Redefining $p = \frac{2\pi n}{L} - \chi$ as $-p$, the entire set
of the quantized momenta can be written as
\begin{equation}
 p = \frac{2\pi n}{L} + \chi,
\end{equation}
where $-\infty < p < \infty$.

In the presence of the mass, there is a potential bound state
solution, which corresponds to an imaginary momentum~\eqref{eq.p_imag}.
If we consider the $L \to \infty$ limit, the bound state
wavefunction should read
\begin{equation}
 e^{- \lambda^* |x|},
\end{equation}
where $x<0$ corresponds to $L-x$.
This corresponds to the situation where only the
``outgoing'' plane waves, $e^{i p^* x}$ for $x>0$,
and $e^{- i p^* x}$ for $x<0$, exist around the defect,
with the imaginary momentum
\begin{equation}
p^* = i \lambda^*.
\label{eq.pstar_imag}
\end{equation}
In an infinite size system $L \to \infty$,
this requires divergence of transmission and reflection
coefficients~\eqref{eq.Tcoeff},~\eqref{eq.RRcoeff},
and~\eqref{eq.RLcoeff}.
Thus
\begin{equation}
 \lambda^* = - m \sin{\chi} > 0.
\label{eq.lambda_Linfty}
\end{equation}
This implies that, when $0< \chi < \pi$, which correspond to
$|b|< 1$, the
bound state exists if and only if $m <0$,
and when $\pi < \chi < 3\pi/2$  or $0 > \chi > -\pi/2$, which
correspond to $|b|>1$, bound state exists if and only if $m>0$.
Although $\chi$ is replaced by $\tilde{\chi}$ in the odd fermion
number sector, the condition for the existence of the bound
state at the defect in an infinite system remains the same,
as $\sin{\tilde{\chi}} = \sin{\chi}$.

The condition for the appearance of the bound state
may be ``physically'' understood as follows.
The ``weak'' defect with $|b|<1$ corresponds to a short
(two-site) section of the Ising chain in the disordered phase,
or equivalently in the trivial (non-topological) phase
in the context of the Kitaev chain.
Similarly, the ``strong'' defect with $|b|>1$ corresponds
to a two-site section of the Kitaev chain in the topological phase.
The bound states appear (only) between the regions
in the trivial and topological phases of the Kitaev chain;
it is the case when the bulk is in the disordered phase ($m>0$)
and the defect is weak ($|b|<1$), or the bulk is in the
ordered phase ($m<0$) and the defect is strong ($|b|>1$).

However, it should be noted that, for the general strength
of the defect, the bound state does \emph{not}
correspond to Majorana zero modes.
This is because, even when the bulk is infinite, the
bound states appearing at the boundaries of the topological
and trivial regions of the Kitaev chain ``talk to each other''
through the short section correspond to the defect.
In fact, from Eq.~\eqref{eq.lambda_Linfty}
the bound-state energy in the thermodynamic limit reads
\begin{equation}
 \epsilon_b = \sqrt{-{\lambda^*}^2+m^2} = |m \cos{\chi} | .
\label{eq.bsenergy_chi}
\end{equation}
When $m \neq 0$, this vanishes only for
the zero defect coupling $b=0$ which correspond
to the open boundary condition (discussed in Sec.~\ref{sec.open})
or the infinitely strong defect $b=\pm \infty$ (discussed below).
The finite bound-state energy~\eqref{eq.bsenergy_chi} implies that
the asymptotic ground-degeneracy in the thermodynamic limit
in the ordered phase, which is physically expected, cannot
be accounted for the generic defects only by the bound-state
formation. As I will discuss below, the other necessary mechanism
is the effective reversal of the fermion number parity as discussed
for the periodic boundary condition in Sec.~\ref{sec.PBC}.
In a more general setting, the bound-state energy vanishes
in the thermodynamic limit under a certain condition
for a generalized defect in
the Kitaev chain
(see Ref.~\cite{KKWK2017} and Appendix~\ref{app.Kitaev_scatt}).

In a finite-size system,
the condition for the existence of
the ``bound state'' with energy smaller than $|m|$
becomes more complicated and 
does depend on the sector, as shown below.
Replacing the momentum $p$ in Eq.~\eqref{eq.fp_def} by the
imaginary momentum~\eqref{eq.p_imag}, I obtain the equation
\begin{equation}
 g (\lambda, \chi) = 0,
\label{eq.glambda_0}
\end{equation}
where 
\begin{align}
 g(\lambda,\chi) & \equiv
(\lambda + m \sin{\chi}) e^{\lambda L}
- 2  \lambda \cos{\chi}
+ e^{- 2 \lambda L} (\lambda - m \sin{\chi}) .
\\
&=
2 \left(
\lambda \cosh{\lambda L}  -  \lambda \cos{\chi}
+  m \sinh{\lambda L}
 \right)
\end{align}
Eq.~\eqref{eq.glambda_0} has a trivial (and unphysical)
solution $\lambda=0$, which corresponds to $p=0$.
The possible bound state is represented by a nontrivial solution
$\lambda >0$ of Eq.~\eqref{eq.glambda_0}.
As in the case of $f(\lambda,\chi)$,
since $g(\lambda,\chi)$ is an odd function of $\lambda$,
its roots appear in pairs $\lambda, - \lambda$.
We can formally include the root $-\lambda$ and
then account for the double-counting.

Obviously,
\begin{align}
g(\lambda,\chi) \sim \lambda e^{\lambda L} \to +\infty
& (\lambda \to \infty) ,
\end{align}
On the other hand,
\begin{equation}
g(\lambda,\chi) \sim 2 \left(1 - \cos{\chi} +  L m \sin{\chi} \right) \lambda
+ O(\lambda^2).
\end{equation}
Thus, there exists a real root $\lambda$ (corresponding to a ``bound state'')
if
\begin{equation}
 2 \sin^2{\frac{\chi}{2}} + 2 L m \sin{\frac{\chi}{2}} \cos{\frac{\chi}{2}}
 = 2 \sin^2{\frac{\chi}{2}} (1 + L m \cot{\frac{\chi}{2}}) < 0,
\end{equation}
namely
\begin{equation}
 mL \cot{\frac{\chi}{2}} < -1 .
\label{eq.bstate_cond}
\end{equation}
This implies that, the bound state exists for a sufficiently large $L$
for $m<0$ if $0<\chi<\pi$, and
for $m>0$ if $-\pi/2 < \chi <0$ or $\pi < \chi < 3 \pi/2$.
In particular, in the limit $L \to \infty$, $e^{-\lambda L} \to 0$ for any
$\lambda >0$, and thus we recover Eq.~\eqref{eq.lambda_Linfty}
from Eq.~\eqref{eq.glambda_0}.

Because the energy eigenvalue $\pm \sqrt{\lambda^2+m^2}$
of any eigenstate in the finite ring
of length $L$ must be real, a real root $\lambda^*$ of
$g(\lambda,\chi)$ (corresponding to a purely imaginary momentum)
has to satisfy
\begin{equation}
 | \lambda^* | \leq |m| .
\end{equation}
As we will see, it is important to know whether and when
the root $\lambda^*$ hits $|m|$, or equivalently when
the energy eigenvalue becomes exactly zero.
Assuming  that there is a root $\lambda^*=m$, in the even sector
with the phase shift parameter $\chi$,
\begin{align}
e^{2 \lambda^* L} g(\lambda^*, L) &=
m ( 1 + \sin{\chi}) e^{2 m L}
- 2 m \cos{\chi} e^{m L}
+ m (1 - \sin{\chi}) 
\\
& =
m  (1 + \sin{\chi})
\left(
e^{ m L} - \frac{\cos{\chi}}{1 + \sin{\chi}}
\right)^2 .
\end{align}
Thus there is a double root $\lambda^* = m$ when
\begin{equation}
e^{ m L} = \frac{\cos{\chi}}{1 + \sin{\chi}},
\end{equation}
and $\lambda^* = -m$ when
\begin{equation}
e^{ - m L} = \frac{\cos{\chi}}{1 - \sin{\chi}}.
\end{equation}
These roots appear when $\cos{\chi}>0$,
namely $-\pi/2 < \chi < \pi/2$.
Within this region, if $-\pi/2 < \chi < 0$,
$\sin{\chi}<0$ and thus $m > 0$ for this ``touching'' to occur.
Likewise, when $0 < \chi < \pi/2$, $m<0$ is required for the touching.
Let us define
\begin{equation}
 s_*(\chi)
\end{equation}
as the solution of
\begin{equation}
 e^{| s_*(\chi)|} = \frac{\cos{\chi}}{1 \pm \sin{\chi}} .
\end{equation}
At the touching point, the bound-state energy is exactly zero
in the finite-size chain, corresponding to exact Majorana
zero modes.
Such a phenomenon is studied in the context of a finite Kitaev ring
with a magnetic flux piercing through the ring
in Ref.~\cite{Navaetal-Majorana2017}.
However, in the present context of the Ising chain,
the touching occurs at different points in the different
fermion-number parity sectors because of the different boundary
conditions.

When the parameter $s=mL$ is varied, the energy of the bound state
crosses zero when the touching occurs at $s=s_*(\chi)$ in the
even sector, and at $s=mL=s_*(\tilde{\chi})$ in the odd sector.
That is, the energy of the bound state changes its sign
between $+\sqrt{{\lambda^*}^2+m^2}$
and $-\sqrt{{\lambda^*}^2+m^2}$ at $s=mL=s_*(\chi)$.
This is very similar to what happens to the $p=0$ eigenstate
when the phase shift $\tilde{\chi}$ is zero in the odd fermion
number sector for the periodic boundary condition,
as discussed in Sec.~\ref{sec.PBC}.
A negative energy eigenvalue of a single-particle eigenstate can be
always made positive by the particle-hole transformation.
However, the particle-hole transformation on a single-particle
eigenstate effectively flips the fermion number parity,
as in Sec.~\ref{sec.PBC}.

Let us call the case where all the energy eigenvalues are positive
without a fermion number parity flip as ``regular'', and
those with a fermion number parity flip as ``reversed''.
Which side of the touching $\lambda^* = |m|$ corresponds to regular
or reversed case can be determined by the continuity from
the periodic boundary condition.
That is, deep inside the disordered phase $ m L \to \infty$
for the general boundary condition $-\infty < b < \infty$,
the situation should be regular and there must be indeed even or odd
number of \textit{positive-energy} fermions in the even and
odd sectors, respectively.
For a given boundary condition $b$ and a fixed $L$,
consider reducing $m$ from $\infty$.
When $mL$ crosses the touching point $s_*(\chi)$,
the even fermion number sector becomes reversed for $mL < s_*(\chi)$.
Likewise, the odd fermion number sector
is reversed for $mL < s_*(\tilde{\chi})$.

\begin{figure}
\begin{center}
\includegraphics[width=0.8\textwidth]{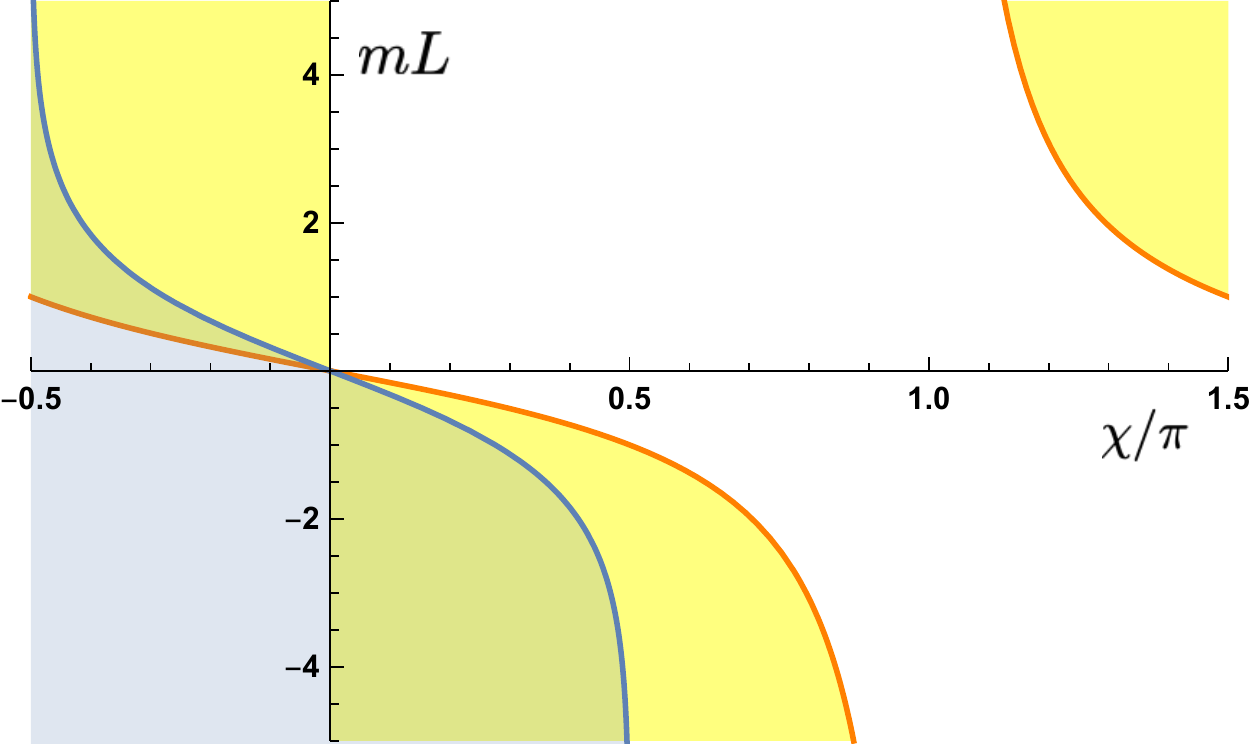}
\includegraphics[width=0.8\textwidth]{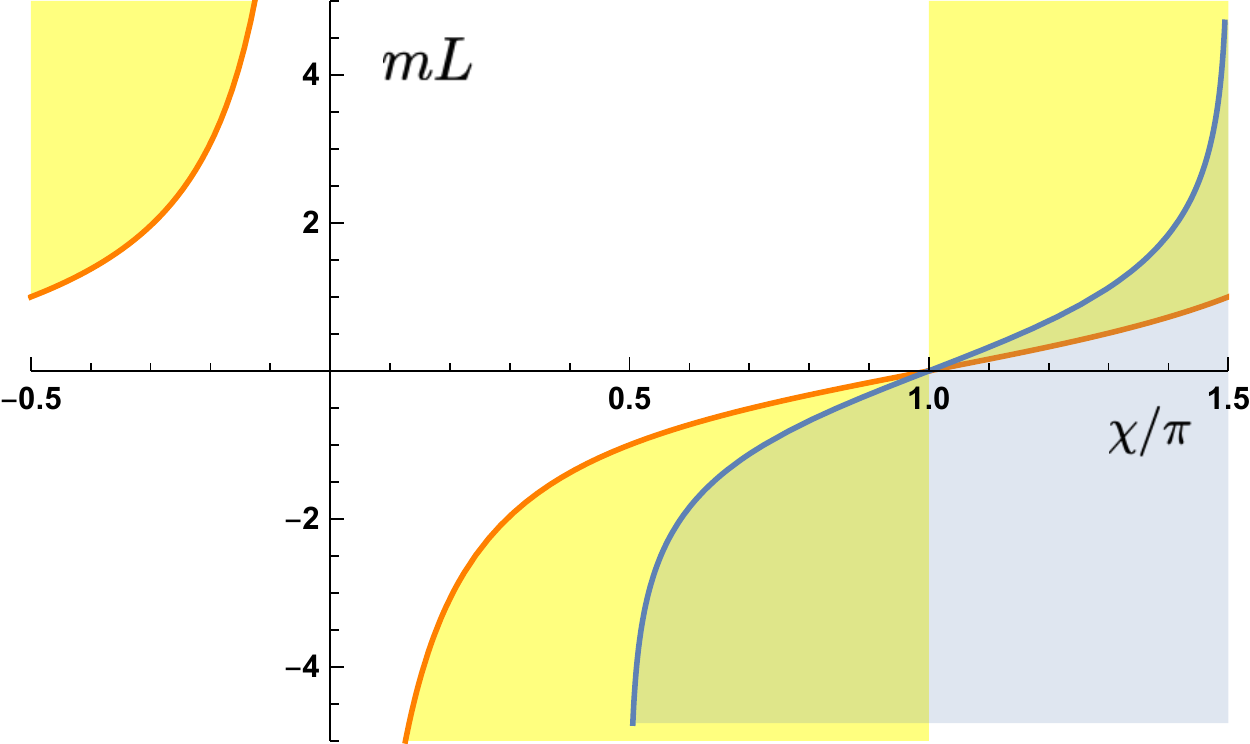}
\caption{
(a)The finite-size ``phase diagram'' for the even fermion
number sector, in the two-dimensional parameter space
$(\chi/\pi,mL)$.
A bound state exists around the defect in the yellow-colored
region. In the blue-colored region, the fermion number parity
is effectively reversed to odd, and there must be at least one
positive-energy particle in the ground state of this sector.
(b) The same ``phase diagram'' for the odd fermion number sector.
If it were drawn with $\tilde{\chi}=\pi-\chi$ as the horizontal
coordinate, it would be identical to (a).
Here, for the sake of the comparison with (a), the same variable
$\chi/\pi$ is used as the horizontal coordinate.
In the blue-colored region, the fermion number parity
is effectively reversed to even, so that the ground state
of this sector has no positive-energy particle.
}
\label{fig.phase_diag}
\end{center}
\end{figure}

The existence of the bound state, and whether the fermion
number parity is reversed, are summarized in Table~\ref{tab.chi_b}.
They are also shown as ``phase diagrams'' in Fig.~\ref{fig.phase_diag}.

I define $p^*(\chi)$ as the ``minimum root''
(smallest positive root) of Eq.~\eqref{eq.fpchi0},
when Eq.~\eqref{eq.bstate_cond} does not hold.
When Eq.~\eqref{eq.bstate_cond} holds,
$p^*(\chi) = i \lambda^*(\chi)$ where $\lambda^*(\chi)$
is the positive (real) root of Eq.~\eqref{eq.glambda_0}.
Then the contribution of the occupied state energy to the
lowest gap is
\begin{align}
 E^{\mbox{\scriptsize diff}}_{\mbox{\scriptsize occ}}
& =
\Theta\left( mL-s_*(\tilde{\chi}) \right) \sqrt{{p^*(\tilde{\chi})}^2+m^2}
- \Theta\left( s_*(\chi)-mL \right) \sqrt{{p^*(\chi)}^2+m^2} .
\label{eq.gap_occ}
\end{align}

Now let us consider the zero-point energies of the ``vacua''
(without any positive-energy fermion)
in Eq.~\eqref{eq.qIsingHcont2} for the general boundary conditions.
In the even fermion number sector, it reads
\begin{equation}
 E^{\mbox{\scriptsize even}}_{\mbox{\scriptsize zero}} =
- \frac{1}{2} \sum_{p>0 | f(p,\chi)=0} \sqrt{p^2+m^2} .
\label{eq.Ezero}
\end{equation}
By construction, the physical eigenstates are characterized by
the positive solutions $p>0$ 
and a possible bound state solution $p= i \lambda$ with $\lambda>0$,
of Eq.~\eqref{eq.fpchi0}.
However, Eq.~\eqref{eq.fpchi0} itself has other solutions.
Indeed, it always has a trivial solution $p=0$, which
is unphysical (corresponding to no actual eigenstate).
Moreover, there are negative roots $p<0$ of Eq.~\eqref{eq.fpchi0} as well.
In fact, since Eq.~\eqref{eq.fp_def} is an odd function of $p$,
all the non-zero roots appear in pairs $-p$ and $p$.
As the pair of the momenta $\pm p$ gives the same energy $\sqrt{p^2+m^2}$,
extending the domain of $p$ to $-\infty < p < \infty$
simply double counts each energy level.
Therefore, Eq.~\eqref{eq.Ezero} can be written as
\begin{align}
 E^{\mbox{\scriptsize even}}_{\mbox{\scriptsize zero}} =
- \frac{1}{2} \frac{1}{2} \frac{1}{2\pi i}
\oint_C \frac{f'(p,\chi)}{f(p,\chi)} \sqrt{p^2+m^2} dp,
\end{align}
where one of the factors $1/2$ is to compensate the double counting
and 
The zero-point energy in the odd sector
$E^{\mbox{\scriptsize odd}}_{\mbox{\scriptsize zero}}$
is then simply given by replacing $\chi$ by $\tilde{\chi}$ in the
above.
There are common contributions from the spurious solution $p=0$
in both sectors, which cancel out in the difference.
The difference can be written as a contour integral
\begin{align}
 E^{\mbox{\scriptsize diff}}_{\mbox{\scriptsize zero}} & =
 E^{\mbox{\scriptsize odd}}_{\mbox{\scriptsize zero}} -
 E^{\mbox{\scriptsize even}}_{\mbox{\scriptsize zero}} \nonumber \\
&= 
\oint_C
\left[
\frac{f'(p,\chi)}{f(p,\chi)} -
\frac{f'(p,\tilde{\chi})}{f(p,\tilde{\chi})}
\right]
\sqrt{p^2+m^2} dp,
\end{align}
where the contour is taken as in Fig.~\ref{fig.contour1}.

Deforming the contour as in Fig.~\ref{fig.contour2}, 
and adding the occupied state energy contribution~\eqref{eq.gap_occ},
I find the universal scaling function for the lowest gap
\begin{align}
\epsmin  =& 
\Theta\left( mL-s_*(\tilde{\chi}) \right) \sqrt{{p^*(\tilde{\chi})}^2+m^2}
- \Theta\left( s_*(\chi)-mL \right) \sqrt{{p^*(\chi)}^2+m^2}
\notag \\
& - \frac{4 }{\pi }
\int_{|m|}^\infty d \kappa \;
\frac{
\left[
\cos{\chi} e^{2 \kappa L}
\left(\sinh{\kappa L} \left(\kappa^2 L-m \sin{\chi} \right)+
\kappa L m \sin{\chi} \cosh{\kappa L}\right)
\right] \sqrt{\kappa^2 - m^2}
}{
e^{4 \kappa L} (m \sin{\chi}+\kappa)^2
-e^{2 \kappa L} \left[
\cos{2 \chi} \left( 2 \kappa^2-m^2\right)+m^2\right]
+ (\kappa - m \sin{\chi})^2
}  .
\label{eq.epsmin_final}
\end{align}
In terms of the scaling variables,
\begin{align}
\Delta & =
\Theta\left( s -s_*(\tilde{\chi}) \right) \sqrt{{q^*(\tilde{\chi})}^2+s^2}
- \Theta\left( s_*(\chi)- s \right) \sqrt{{q^*(\chi)}^2+m^2}
\notag \\
& - \frac{4 }{\pi }
\int_{|s|}^\infty d y \;
\frac{
\left[
\cos{\chi} e^{2 y}
\left(\sinh{y} \left(y^2 - s \sin{\chi} \right)+
y s \sin{\chi} \cosh{y}\right)
\right] \sqrt{y^2 - s^2}
}{
e^{4 y} (s \sin{\chi}+ y)^2
-e^{2 y } \left[
\cos{2 \chi} \left( 2 y^2- s^2\right)+ s^2\right]
+ (y - s \sin{\chi})^2
}  .,
\label{eq.Delta_final}
\end{align}
where $q^*(\chi)$ is the smallest positive root of
\begin{equation}
 q \cos{q} - q \cos{\chi} + s \sin{\chi} \sin{q} =0,
\end{equation}
or its pure imaginary root when $ s \cot{\frac{\chi}{2}} < -1$.

Below I will remark several observations.
In the thermodynamic limit $L \to \infty$
for any fixed non-zero mass $m \neq 0$,
the difference in the zero-point energies
$E^{\mbox{\scriptsize diff}}_{\mbox{\scriptsize zero}}$
vanishes exponentially
(See also Ref.~\cite{AbrahamKoSvrakic1989}).
Thus the lowest gap is determined by
the occupied state contribution~\eqref{eq.gap_occ}.

In the thermodynamic limit in the disordered phase, $s=mL \to +\infty$,
there is no reversal of the fermion number parity in either sector.
As a consequence, the lowest gap in this case is determined by
the lowest single-particle energy in the odd sector:
\begin{equation}
 \epsmin \sim \sqrt{\left(p^*(\tilde{\chi})\right)^2+m^2} .
\end{equation}
When the defect (boundary condition) is ``weak''
($|b| \leq 1$), there is no bound state in this limit, resulting in
\begin{equation}
 \epsmin \sim m.
\end{equation}
On the other hand, when the defect is ``strong'' ($|b|>1$),
a bound state localized near the defect appears in the disordered
phase.
In the thermodynamic limit, the localized state is given by
Eqs.~\eqref{eq.pstar_imag} and~\eqref{eq.lambda_Linfty},
and I find
\begin{equation}
 \epsmin \sim  m |\cos{\chi}| .
\end{equation}

As the mass parameter $m$ is decreased, a bound state in either
sector may go under the reversal (crosses zero energy).
The results on the existence of the bound state and the
fermion number parity reversal, in the range
of the parameter $b$, is summarized in Table~\ref{tab.chi_b}. 
I find that, for a general value of parameter $b$ except for
$b = 0$ or $b=\pm \infty$, the reversal occurs exactly once
in one of the sectors (even or odd fermion number)
in the range $-\infty < m < \infty$.
This means that, deep inside the ordered phase $mL \to -\infty$,
both sectors actually have the same effective fermion number
parity defined with respect to the positive energy states.
This results in the asymptotic degeneracy of the ground states
\begin{equation}
 \epsmin \to 0 ,
\end{equation}
in the thermodynamic limit in the ordered phase for any
boundary condition. 
As discussed earlier, this is physically a consequence of the
spontaneous breaking of the $\mathbb{Z}_2$ symmetry in the
thermodynamic limit of the ordered phase $m<0$.

As the defect strength approaches infinity ($|b| \to \infty$),
the range of $mL$ in which the reversal occurs in either sector
grows.
In the limit $b = \pm \infty$,
$\chi = - \pi/2$ or $3\pi/2$,
the reversal occurs in one of the sectors
for any value of $s=mL$.
Furthermore, in this limit, the phase shifts in two sectors are equivalent
($\chi \equiv \tilde{\chi} \mod{2\pi}$), resulting in an exact two-fold
degeneracy of the entire energy spectrum for any value of $s=mL$.
This exact degeneracy can be simply understood as follows.
In the limit $b \to \pm \infty$, the two spins across the defect
are strongly coupled and thus can be replaced by a single effective spin.
For example, in the infinitely strong ferromagnetic limit $b=\infty$,
the effective spin at the defect represents one of the two states
$\uparrow \uparrow$ and $\downarrow \downarrow$.
It is important to observe that two spin flips are required to
transit between these two states, with the intermediate state
of an energy higher than the two states by $\sim b$.
Thus, in the limit $b=\infty$, the transverse field for the effective
spin vanishes, and the system can be described by the effective model
\begin{equation}
\calH_I = - J \sum_{j=1}^{N-1}  \sigma^z_j \sigma^z_{j+1}
- \Gamma \sum_{j=1}^{N-2} \sigma^x_j ,
\label{eq.qIsing_binfty}
\end{equation}
where $j=N-1$ corresponds to the effective spin representing
the two strongly coupled spins $j=N-1$ and $j=N$ in the original
model~\eqref{eq.qIsing_b}, and
$\sigma^\alpha_N \equiv \sigma^\alpha_1$.
Since there is no transverse field for the effective spin at $j=N-1$,
$\sigma^z_{N-1}$ commutes with the Hamiltonian and is thus conserved.
As a consequence, every eigenstate is exactly two-fold degenerate,
corresponding to $\sigma^z_{N-1}=\pm 1$.
Similar argument applies to the infinitely strong antiferromagnetic
defect $b=-\infty$.
I note that, for the defect line in
the classical 2-dimensional Ising model with a finite ratio
of couplings,
these limits are never reached and the exact degeneracy of
the spectrum of the transfer matrix is not realized.

\section{Conclusion}
\label{sec.conclusion}

In this paper, I derived the universal finite-size scaling function
of the lowest gap in the quantum Ising chain in the scaling limit
around the quantum critical point, with the general ``defect''
boundary conditions.
Although the derivation was based on the exact solution of the
quantum Ising chain, the same scaling function should apply
to more generic, non-integrable models in the scaling limit,
as long as they belong to the same Ising universality class.
It should be noted, however, that corrections from irrelevant
operator (in the renormalization group sense) do exist,
and might be important in fitting numerical/experimental data.
(The scaling limit discussed in this paper is the limit where
the corrections from irrelevant operators can be ignored).
These corrections are outside the scope of the present paper.

The most interesting feature is the asymmetry of the scaling function
with respect to the mass inversion (Kramers-Wannier duality):
even though the bulk dispersion relation of the relativistic
Majorana fermion field theory appears to be symmetric,
the lowest gap asymptotically vanishes in the thermodynamic
limit in the ordered phase (negative mass in our convention).
This asymptotic ground-state degeneracy (vanishing of the lowest
gap in the thermodynamic limit) is a consequence of the
spontaneous symmetry breaking in the ordered phase.
Since the ``defect'' boundary condition discussed in this paper
does not break the $\mathbb{Z}_2$ symmetry of the quantum Ising
chain, the asymptotic ground-state degeneracy in the ordered phase
holds for the entire family of the defect boundary conditions.
While the asymptotic degeneracy in the ordered phase was discussed
for the defect boundary conditions in Ref.~\cite{AbrahamKoSvrakic1989},
the universal scaling function~\eqref{eq.Delta_final}
representing the crossover between the ordered and disordered phases
is elucidated in this paper.
While the analysis in the present paper relies on the exact
solution of the quantum Ising chain, the obtained universal
scaling function should apply to more general class of models,
as long as the bulk quantum critical behavior belongs to the
Ising universality class, and the defect/boundary does not
break the $\mathbb{Z}_2$ symmetry explicitly.
In the absence of the exact solution, however, the parameters
$m$, $\chi$, and the spin-wave velocity, are not a priori known
from the microscopic model, and have to be determined for example
by a numerical fitting.

The asymmetry is caused by the bound state at the defect and by
the effective reversal of the fermion number parity in one
of the sectors.
The bound state occurs in either the ordered or the disordered phase,
depending on the defect strength.
The exceptions are the periodic and antiperiodic boundary conditions,
which do not host a bound state.
For general defect boundary conditions,
the bound states are not exactly at zero energy
in the thermodynamic limit.
However, upon changing the mass parameter, the bound-state energy
in the finite size can hit zero.
The single-particle energy of the bound state would become negative
beyond such a point, if extrapolated analytically.
It is then convenient to apply the particle-hole transformation to the
bound state, which makes the bound-state energy positive.
However, this particle-hole transformation effectively
flips the fermion number parity.
This introduces a Heaviside step function in the universal
scaling function~\eqref{eq.Delta_final}
for the lowest gap, which is a source of the asymmetry between
the ordered and disordered phases.
Although the Heaviside step function itself is non-analytic,
the entire finite-size scaling function is analytic.
The resulting, nontrivial ``phase diagram'' is summarized
in Fig.~\ref{fig.phase_diag} and Table~\ref{tab.chi_b}.

The effective reversal of the fermion number parity occurs
in one of the sectors for general defect boundary conditions,
except for the open boundary condition.
For the open boundary condition, although the fermion number parity
is not reversed in either sector, the bound state in this case
is a pair of the ``Majorana zero modes''.
Thus its energy asymptotically vanishes in the thermodynamic limit.
For other boundary conditions, the effective reversal of the
fermion number parity is necessary to achieve the asymptotic
ground-state degeneracy in the ordered phase.

As a byproduct of the main analysis of the paper, I applied
the present formulation to generalized defects in the Kitaev chain.
The Kitaev chain has more degrees of freedom than what can be
obtained by a mapping from the quantum Ising chain.
In particular, the Kitaev chain admits a more general class
of defects characterized by phase factors.
In Ref.~\cite{KKWK2017}, Majorana zero modes are found in the
thermodynamic limit under the condition,
even when the defect coupling is non-vanishing.
In the present formulation, this corresponds to vanishing
of the transmission amplitude due to an interference,
which makes the defect effectively open boundaries
in the low-energy limit. 
I hope that the present observations will provide a useful
perspective to the finite-size scaling of more general systems,
especially those exhibiting quantum phase transitions.

\section*{Acknowledgements}

This work was initiated around 1995
while I was a Research Associate at University of Tokyo and
then a Killam Post-Doctoral Fellow at University of British Columbia.
I thank Kohei Kawabata for a stimulating comment which
led to the completion of this work, and Hosho Katsura
for useful information on many related literature.
I also acknowledge stimulating discussions with
Ian Affleck, Ryohei Kobayashi,
Subir Sachdev, Kazumitsu Sakai, and Ruben Verresen.
A part of this work were performed at Aspen Center for Physics,
which was supported by the US National Science Foundation
Grant No.PHY-1607611.
This work was supported in part by
MEXT/JSPS KAKENHI Grant Nos. 17H06462 and 19H01808, and
US NSF Grant No. PHY-1748958.

\appendix

\section{Exact Solution of the Ising chain and Majorana fermion field theory}
\label{app.Ising_cont}

For convenience, I rotate the spin axes so that the quantum Ising chain
Hamiltonian reads
\begin{equation}
\calH_I = - J \sum_j  \sigma^x_j \sigma^x_{j+1}
- \Gamma \sum_j \sigma^z_j .
\label{eq.qIsingH2}
\end{equation}
The Jordan-Wigner transformation is given as
\begin{align}
  \sigma^+_j & = 2 \exp{\left( i \pi \sum_{l<j} n_l \right)}c^\dagger_j  ,
\\
  \sigma^-_j & = 2 \exp{\left( i \pi \sum_{l<j} n_l \right)}c_j ,
\\
  \sigma^z_j &= 2 n_j - 1
\end{align}
where $c_j$ and $c^\dagger_j$ are standard fermion
annihilation/creation operators, and
\begin{equation}
 n_j = c^\dagger_j c_j ,
\end{equation}
is the fermion number operator.
By the Jordan-Wigner transformation, the quantum Ising chain
Hamiltonian~\eqref{eq.qIsingH2} is transformed to
the Kitaev chain~\eqref{eq.Kitaevchain}, with the parameters.

I first ignore the issue of the boundary condition and
simply apply Fourier transformation to the fermion operators:
\begin{align}
 c_j & = \sqrt{\frac{a}{L}} \sum_k c(k) e^{i k j a}, \\
 c^\dagger_j & = \sqrt{\frac{a}{L}} \sum_k c^\dagger(k) e^{- i k j a}, \\
\end{align}
where $a$ is the lattice constant and $L$ is the length of the chain.
The Kitaev chain Hamiltonian can then be rewritten as
\begin{align}
 \calH_K = \frac{1}{2} \sum_k
\begin{pmatrix}
  c^\dagger(k) & c(-k)
\end{pmatrix}
\hBdG (k)
\begin{pmatrix}
  c(k) \\ c^\dagger(-k)
\end{pmatrix} .
\end{align}
Here the ``Bogoliubov-de Gennes'' Hamiltonian reads
\begin{equation}
\hBdG(k) \equiv (-2t \cos{(k a)} - \mu ) \tau_z
+ 2 \Delta \sin{(k a)} \tau_y,
\end{equation}
where $\tau_\alpha$ are Pauli matrices acting on the Nambu space.
For $\Delta \neq 0$, the energy gap closes when
\begin{equation}
 2 |t | = |\mu |,
\end{equation}
at either $k=0$ or $k=\pi/a$.
For the choice of parameters $J >0, \Gamma>0$, the gap closes
at $k= \pi/a$ when $J = \Gamma$.
Since we are interested in the universal behavior in the vicinity
of this quantum critical point, we set
\begin{align}
 J &= \frac{1}{2a} ,
\label{eq.Jcont}
 \\
 \Gamma &= \frac{1}{2a} + \frac{m}{2},
\label{eq.Gammacont}
\\
 p & = k - \frac{\pi}{a} ,
\label{eq.pcont}
\end{align}
and consider the continuum (scaling) limit $a \to 0$.
In the context of the Kitaev chain, Eqs.~\eqref{eq.Jcont}
and \eqref{eq.Gammacont} implies
\begin{align}
 t = \Delta & = \frac{1}{2a},
\label{eq.tDeltacont}
\\
 \mu &= \frac{1}{a} + m .
\label{eq.mucont}
\end{align}

The resulting Bogoliubov-de Gennes Hamiltonian is
\begin{equation}
\thBdG(p) \sim - m \tau^z - p \tau^y .
\label{eq.hBdG_p}
\end{equation}
Fourier transforming back, it reads
\begin{equation}
\thBdG(x) \sim - m \tau^z + i \tau^y \partial_x ,
\label{eq.hBdGcont}
\end{equation}
in the real space.
The entire many-body Hamiltonian in the continuum limit is
\begin{equation}
 \calH_I \sim \frac{1}{2} \int dx\; \Bpsi^\dagger(x) \thBdG(x) \Bpsi(x),
\label{eq.qIsingcont}
\end{equation}
where
\begin{equation}
 \Bpsi(x) = (-1)^{x/a}
\begin{pmatrix}
 c(x) \\
 c^\dagger(x) 
\end{pmatrix}  
.
\end{equation}
The single-particle Hamiltonian~\eqref{eq.hBdG_p}
has the energy eigenvalues $\pm \epsilon(p)$ and corresponding
eigenvectors
\begin{align}
 \thBdG(p) \tBphi_\pm(p) = \pm \epsilon(p) \tBphi_\pm(p),
\end{align}
where $\epsilon(p)$ is defined in Eq.~\eqref{eq.dispersion}
and double sign corresponds.
Explicitly,
\begin{align}
 \tBphi_+(p) &= \frac{1}{\sqrt{2 (p^2+m^2+\sqrt{m^2+p^2})}}
\begin{pmatrix}
 i p \\
 m + \sqrt{p^2+m^2}
\end{pmatrix} ,
\label{eq.tBphi_p}
\\
 \tBphi_-(p) &= \frac{1}{\sqrt{2 (p^2+m^2+\sqrt{m^2+p^2})}}
\begin{pmatrix}
 m + \sqrt{p^2+m^2} \\
 i p
\end{pmatrix}.
\end{align}

The many-body Hamiltonian~\eqref{eq.qIsingcont} is then diagonalized as 
\begin{equation}
 \calH_I \sim \frac{1}{2} \sum_p \epsilon(p)
  \left( {\eta^{(+)}}^\dagger(p) \eta^{(+)}(p) -
 {\eta^{(-)}}^\dagger(p) \eta^{(-)}(p)
\right),
\label{eq.qIsing_2bands}
\end{equation}
where
\begin{align}
\eta^{(\pm)}(p)  &= \tphi_\pm(p)^\dagger 
\begin{pmatrix}
 c(p) \\
 c^\dagger(-p) 
\end{pmatrix}  
.
\end{align} 

However, it should be noted that $\thBdG$ was obtained after
formally doubling the degrees of freedom,
by regarding $c$ and $c^\dagger$ as independent operators, although
they are actually not.
Reflecting this, $\hBdG$ has an enforced particle-hole
(charge conjugation) symmetry.
The particle-hole transformation $\calC$ on the many-body
Hilbert space is characterized by
\begin{align}
 \calC^{-1} \alpha c(x) \calC & = \alpha^* c^\dagger(x), \\
 \calC^{-1} \alpha^* c^\dagger(x) \calC & = \alpha c(x) ,
\end{align}
where $\alpha$ is a complex number (scalar).
The particle-hole symmetry of the system implies
\begin{equation}
 \calC^{-1} \calH_I \calC = - \calH_I .
\end{equation}
In the representation~\eqref{eq.hBdGcont},
\begin{equation}
 \calC^{-1} \calH_I \calC =
  \frac{1}{2} \int dx\; \Psi^\dagger(x)
  \left[ \tXi^{-1} \thBdG(x) \tXi \right] \Psi(x),
\end{equation}
where $\tXi$ is the particle-hole transformation acting on the
spinor (single-particle) states
\begin{equation}
 \tXi \equiv \tau^x \mathcal{K},
\end{equation}
with $\mathcal{K}$ denotes the complex conjugation.

The particle-hole symmetry of the many-body Hamiltonian is
then reduced to that of the single-particle one: 
\begin{equation}
 \tXi \thBdG \tXi^{-1} = - \thBdG .
\end{equation}
As a consequence, for any eigenstate of $\thBdG$ with a non-vanishing
energy $\epsilon$, there is a ``particle-hole mirror'' eigenstate
with the energy $-\epsilon$.
That is, if
\begin{equation}
 \thBdG \tBphi_\epsilon = \epsilon \tBphi_\epsilon ,
\end{equation}
\begin{equation}
 \thBdG \left( \tXi \tBphi_\epsilon \right)
= - \epsilon \left( \tXi \tBphi_\epsilon \right) .
\end{equation}
In fact, by construction, these two eigenstates of $\hBdG$ actually
refer to the same physical degree of freedom.
In other words, we can identify the creation operator for the
eigenstate with energy $\epsilon$ with the annihilation operator
for the particle-hole mirror:
\begin{equation}
 \eta^\dagger_\epsilon = \eta_{-\epsilon} .
\end{equation}
As an exception, an eigenstate with exact zero energy eigenvalue is
the particle-hole mirror of itself (Majorana zero mode).

In fact, we can explicitly see that
\begin{equation}
 \eta_+(p) = \left\{ \eta_-(-p) \right\}^\dagger  .
\end{equation}
This means that, although it might appear that the effective
Hamiltonian~\eqref{eq.qIsing_2bands} has ``2 bands'', in fact
there is only one band (one eigenmode for each momentum),
and the Hamiltonian may be written as
\begin{equation}
 \calH_I \sim \sum_p \epsilon(p)
  \left( \eta^\dagger(p) \eta(p) - \frac{1}{2}
\right) ,
\label{eq.qIsingHcont2}
\end{equation}
where $\eta(p)=\eta^{(+)}(p)$.

On the other hand, the relativistic Dirac fermion
in $d+1$-dimensional space-time is defined by
the Lagrangian density
\begin{equation}
 \calL_D = \bar{\Bpsi} \left( i \gamma^\mu \partial_\mu - m \right) \Bpsi ,
\end{equation}
where the summation over $\mu=0,1,\ldots,d$ is implicitly assumed.
$\gamma^\mu$'s are Dirac $\gamma$-matrices satisfying
\begin{equation}
 \left\{  \gamma^\mu, \gamma^\nu \right\} = 2 \eta^{\mu \nu},
\end{equation}
where $\eta^{\mu \nu}$ is the Minkowski metric
$\eta^{00}= - \eta^{ii}= 1$ ($i=1,\ldots,d$).
The (minimal) dimension of $\gamma$-matrices is known to be
\begin{equation}
2^{\lfloor (d+1)/2 \rfloor} .
\end{equation}
In particular, for one spatial dimension I am going to discuss
in this paper, $\gamma$-matrices are $2 \times 2$ matrices.

The Hamiltonian density can be obtained by the canonical quantization as
\begin{equation}
 \calH_D = \psi^\dagger
\left( - i \alpha^j \partial_j + m \gamma^0 \right) \psi,
\end{equation}
where the summation over $j=1,\ldots, d$ is implicitly assumed and
\begin{align}
 \Bpsi^\dagger & \equiv \bar{\Bpsi} \gamma^0, \\
 \alpha^j & \equiv \gamma^0 \gamma^j . 
\end{align}
By definition
\begin{equation}
 \left\{  \alpha^i , \alpha^j \right\} = 2 \delta^{ij} .
\end{equation}
Identifying $\gamma^0 = - \tau^z$ and $\alpha^1 = - \tau^y$,
Eq.~\eqref{eq.hBdGcont} is nothing but the Dirac Hamiltonian
in 1+1 dimensions.

At the critical point $\Gamma = \Gamma_c$, in the continuum (scaling)
limit we have relativistic massless fermions $m=0$.
In this case, Eq.~\eqref{eq.hBdGcont} reduces to
\begin{equation}
\thBdG(p) \sim  - p \tau^y ,
\label{eq.hBdGmassless}
\end{equation}
with the eigenstates
\begin{align}
 \tBphi_R(p) & = \frac{1}{\sqrt{2}}
\begin{pmatrix}
i \\
1				
\end{pmatrix},
\\
 \tBphi_L(p) & = \frac{1}{\sqrt{2}}
\begin{pmatrix}
1 \\
i				
\end{pmatrix}.
\end{align}
Thus, defining
\begin{equation}
U \equiv
\frac{1}{2}
\begin{pmatrix}
 1 + i & 1 - i \\
 1 - i & 1 + i
\end{pmatrix},
\end{equation}
Eq.~\eqref{eq.hBdGmassless} is transformed to
\begin{equation}
 U^\dagger \left( -p \tau^y \right) U = p \tau^z .
\end{equation}
This corresponds to ``chiral basis'' in which the
spinor field can be written as
\begin{equation}
 \Bpsi =
\begin{pmatrix}
 \psi_R \\ \psi_L
\end{pmatrix}  ,
\end{equation}
in terms of the right-mover $\psi_R$ and $\psi_L$, with
the dispersion $\epsilon = \pm p$ respectively.
More generally, including the  mass term $m$,
the effective Hamiltonian
in the chiral basis is
\begin{equation}
 U^\dagger \hBdG(p) U = p \tau^z - m \tau^y .
\end{equation}
In this chiral basis, the particle-hole transformation reads
\begin{equation}
 \Xi =  U^\dagger \tau^x \mathcal{K} U
 = \mathcal{K} .
\end{equation}
The imposed particle-hole symmetry thus means that the
fields $\psi_{R,L}$ are real (Majorana) fields.

\section{Scattering at the Ising defect in the scaling limit}
\label{app.scatt_defect}

Let us consider the Kitaev chain with a defect between sites
$j=0$ and $j=1$.
\begin{equation}
 \calH_d =  \frakb \left\{
- t \left( c^\dagger_{0} c_1 + c^\dagger_1 c_{0} \right)
+ \Delta \left( c_{0} c_1 + c^\dagger_1 c^\dagger_0 \right)
\right\},
\end{equation}
which is obtained from the Ising chain
with the defect~\eqref{eq.qIsing_b}.

First we solve Kitaev chain in the bulk and in the continuum
limit, in an alternative manner.
A single-particle eigenstate of the BdG Hamiltonian in the bulk
satisfies
\begin{align}
\begin{pmatrix}
 -t & \Delta
\\
-\Delta & t 
\end{pmatrix} 
\tBphi_{j-1}
-
\begin{pmatrix}
 - \mu - \epsilon & 0 \\
0 & \mu - \epsilon
\end{pmatrix}
\tBphi_j
+
\begin{pmatrix}
 -t & - \Delta \\
\Delta & t 
\end{pmatrix} 
\tBphi_{j+1}
& =0,
\end{align}
where $(-1)^j \tBphi_j$ is the two-component eigenstate wavefunction
at site $j$. The factor $(-1)^j$ is included for a later convenience,
and corresponds to the $\pi/a$ momentum shift in Eq.~\eqref{eq.pcont}.

For the special case $t=\Delta=J$ which arises from the Ising chain,
a simplification occurs in a new basis.
The wavefunction in the new basis $\ttBphi$ is defined by
\begin{align}
 \tBphi_j &=  V \ttBphi_j,
\label{eq.tBphi_ttBphi}
\end{align}
where
\begin{equation}
 V = \frac{1}{\sqrt{2}}
\begin{pmatrix}
 1 & 1 \\
 1 & -1			 
\end{pmatrix} .
\end{equation}
Namely,
\begin{align}
2J
\begin{pmatrix}
0 & 1
\\
0 & 0
\end{pmatrix} 
\ttBphi_{j-1}
+
\begin{pmatrix}
 - \epsilon & -\mu \\
-\mu &  - \epsilon
\end{pmatrix}
\ttBphi_j
+
2J
\begin{pmatrix}
 0 & 0 \\
 1 & 0
\end{pmatrix} 
\ttBphi_{j+1}
& =0.
\label{eq.latticeSch_tt}
\end{align}
For the choice of the parameters~\eqref{eq.Jcont} and \eqref{eq.mucont},
we find
\begin{align}
 \frac{1}{a}\left( \ttphi^2_{j-1} -  \ttphi^2_j  \right)
-  m \ttphi^2_j
&= \epsilon \ttphi^1_j  \\
 \frac{1}{a}
\left(
\ttphi^1_{j+1} - \ttphi^1_j
\right)
- m \ttphi^2_j
 &= \epsilon \ttphi^2_j .
\end{align}
For a given value of $\epsilon$, these equations give
a recursion relation which determine $\ttBphi_{j+1}$ by
$\ttBphi_j$.
Thus the general eigenstates for a given $\epsilon$
is a two-dimensional (complex) vector space.
This is a special feature which occurs only when $t=\Delta$
in the Kitaev chain.

In the continuum limit, this corresponds to the BdG Hamiltonian
\begin{equation}
 \tthBdG = - i \tau^y \partial_x - m  \tau^x
\end{equation}
and the eigenstates.
The positive energy ($\epsilon = \sqrt{p^2+m^2}$)
plane-wave eigenstates in this basis are given as
\begin{align}
 \ttBphi^+(p)  & \sim
\begin{pmatrix}
   \sqrt{p^2+m^2} \\
 ip -m
\end{pmatrix} .
\label{eq.ttBphi_p}
\end{align}
This is identical to Eq.~\eqref{eq.tBphi_p} via the
transformation~\eqref{eq.tBphi_ttBphi}.
The general eigenstates for $\epsilon=\sqrt{p^2+m^2}$
with $p \neq 0$ are given by
the linear superpositions of
$\ttBphi^+(p)$ and $\ttBphi^+(-p)$.

Around the defect, the eigenequation~\eqref{eq.latticeSch_tt}
is modified as 
\begin{align}
\frac{\frakb}{a} 
\begin{pmatrix}
0 & 1
\\
0 & 0
\end{pmatrix} 
\ttBphi_0
+
\begin{pmatrix}
 - \epsilon & - \frac{1}{a} -m \\
- \frac{1}{a} -m & - \epsilon 
\end{pmatrix}
\ttBphi_1
+
\frac{1}{a}
\begin{pmatrix}
0 & 0 \\
1 & 0
\end{pmatrix} 
\ttBphi_2
& =0,
\\
\frac{1}{a}
\begin{pmatrix}
0 & 1
\\
0 & 0
\end{pmatrix} 
\ttBphi_{-1}
+
\begin{pmatrix}
 - \epsilon & - \frac{1}{a} -m \\
- \frac{1}{a} -m & - \epsilon 
\end{pmatrix}
\ttBphi_0
+
\frac{\frakb}{a}
\begin{pmatrix}
0 & 0 \\
1 & 0
\end{pmatrix} 
\ttBphi_1
& =0 .
\end{align}
In the leading order of $a$, this implies
\begin{align}
\ttphi^2_{-1} &= \ttphi^2_0 ,  
\label{eq.cont1}
\\
\ttphi^1_1 &= \ttphi^1_2 ,
\label{eq.cont2}
\\
\ttphi^1_0 &= \frakb \ttphi^1_1 ,
\label{eq.match1}
\\
\frakb \ttphi^2_0 &=  \ttphi^2_1 .
\label{eq.match2}
\end{align}
In order to obtain the transmission/reflection coefficients,
I take an scattering states ansatz 
\begin{align}
 \ttBphi_j &=
\begin{cases}
 \ttBphi^+(p) e^{i p a j} 
+  R_\rightarrow \ttBphi^+(-p) e^{- i p a j} 
& ( j \leq 0),
\\
 T_\rightarrow \ttBphi^+(p) e^{i p a j} 
& ( j \geq 1),
\end{cases}
\end{align}
with Eq.~\eqref{eq.ttBphi_p},
where $T_\rightarrow$ and $R_\rightarrow$
are transmission/reflection coefficients for the
right-moving incoming fermion with the momentum $p>0$.
In the continuum limit around the defect, $a$ can be simply
set to zero.
Then Eqs.~\eqref{eq.cont1} and~\eqref{eq.cont2}  are automatically
satisfied by the continuity of the wavefunction in either side
of the defect.
In contrast, Eqs.~\eqref{eq.match1} and~\eqref{eq.match2}
give nontrivial constraints on $T_\rightarrow$ and $R_\rightarrow$.
Solving them, we obtain Eqs.~\eqref{eq.Tcoeff} and~\eqref{eq.RRcoeff}, 
with the phase shift parameter $\chi$ as the root of
\begin{equation}
 \frakb =  - \tan{\left[\frac{1}{2}\left(\chi - \frac{\pi}{2}\right)\right]} .
\end{equation}
A similar calculation leads to Eqs.~\eqref{eq.Tcoeff} and~\eqref{eq.RLcoeff}
for $T_\leftarrow$ and $R_\leftarrow$.

The boundary coefficient $\frakb$ of the Kitaev chain depends
on that of the Ising chain $b$ and the fermion number parity.
In the even sector, $\frakb = -b$.
As a consequence, transmission/reflection amplitudes in the
even sector is given as Eqs.~\eqref{eq.Tcoeff},~\eqref{eq.RRcoeff},
and~\eqref{eq.RLcoeff},
with the phase shift parameter Eq.~\eqref{eq.chi_in_b}.
The phase shift parameter in the odd sector, where $\frakb = b$,
should be replaced by $\tilde{\chi}$ as defined in
Eq.~\eqref{eq.def_tildechi}.

\section{Scattering at the Kitaev chain defect}
\label{app.Kitaev_scatt}

If we set aside the quantum Ising chain problem, and
consider the Kitaev chain as the given problem,
we are free from various
constraints originating from the Ising chain.
For example, the constraint $t = \Delta$ no longer exists,
and $t$ and $\Delta$ can be independent complex numbers.
In fact, there is no reason to expect $t=\Delta$ in a physical
realization of the Kitaev chain, except for a fine-tuning.
We can still consider the scaling limit near a critical point
$\mu = \mu_c = 2t$.

In the bulk, the complex phase of $t$ can be eliminated by
a gauge transformation (local phase transformation) of the
complex fermion operator $c$, and the complex phase of $\Delta$
can be eliminated by a phase transformation of Majorana fermion
operators~\cite{Kitaev-chain}.
Thus we assume that $t$ and $\Delta$ in the bulk to be
real and positive.

Still, one can introduce complex phase factors for $t$ and
$\Delta$ at the defect.
The Hamiltonian with such a defect reads
\begin{align}
\calH_K =& 
- t \sum_{j=1}^{N-1} \left( c^\dagger_j c_{j+1} + c^\dagger_{j+1} c_j \right)
- \mu \sum_{j=0}^N c^\dagger_j c_j
+ \sum_{j=1}^{N-1}
\left( \Delta c_j c_{j+1} + \Delta  c^\dagger_{j+1} c^\dagger_j \right),
\nonumber \\
&
+\frakb
\left[
- t ( e^{i \phi_1} c^\dagger_0 c_1 + \mbox{h.c} )
+ \Delta ( e^{i \phi_2} c_j c_{j+1} + \mbox{h.c.} )
\right],
\label{eq.Kitaev_defect}
\end{align}
where $c_N \equiv c_0$.
In fact, this model was studied recently in Ref.~\cite{KKWK2017}, where
Majorana zero modes were found even when the defect coupling $\frakb$
is non-zero, if $\phi_1$ and $\phi_2$ satisfy a certain relation.
It is also interesting to understand this phenomenon in terms of
the effective low-energy field theory.

First we consider the special case $t = \Delta$, which we
encountered in the mapping from the Ising chain, but with
the phase factors $\phi_1, \phi_2$.
In this case, the momentum eigenstates in the bulk is unchanged
from Eq.~\eqref{eq.ttBphi_p}.
On the other hand, matching conditions~\eqref{eq.match2} and~\eqref{eq.match1}
around the defect are modified to
\begin{align}
\ttphi^1_0 & = \frac{\frakb}{2}
\left[
(\cos{\phi_1}+\cos{\phi_2}) \ttphi^1_1
\right]
+ i 
\left[
(\sin{\phi_1}-\sin{\phi_2}) \ttphi^2_1
\right] ,
\\
\ttphi^2_1 &=  \frac{\frakb}{2}
\left[
(\cos{\phi_1}+\cos{\phi_2}) \ttphi^2_0
\right]
- i 
\left[
(\sin{\phi_1} +\sin{\phi_2}) \ttphi^2_0
\right] ,
\end{align}
In the continuum limit.
\begin{align}
 T =& \frac{4 b \sqrt{m^2+p^2} (\cos{\phi_1}+\cos{\phi_2})}{D}
\\
R=&
\frac{1}{D}
\left\{
 \left(b^2-4 \right) \sqrt{m^2+p^2} \right.
\notag \\
& \left.
+ b^2 \left[ 2 \sqrt{m^2+p^2} \cos{\phi_1} \cos{\phi_2}
+\sqrt{m^2+p^2} \cos{2 \phi_2}
-2 \left( \cos{\phi_1} +\cos{\phi_2} \right) 
 \left( p \sin{\phi_1}-i m \sin{\phi_2} \right)
\right]
\right\}
\end{align}
where
\begin{multline}
D = \left(b^2+4\right) \sqrt{m^2+p^2} \\
+b^2 \left[
 \sqrt{m^2+p^2} \cos^2{\phi_1}
+ 2 \sqrt{m^2+p^2} \cos{\phi_1} \cos{\phi_2}
\right. \\
\left.
-\sqrt{m^2+p^2} \sin^2{\phi_1}
+2 (\cos{\phi_1}+\cos{\phi_2}) 
    \left(p \sin{\phi_2} -i m \sin{\phi_1} \right)
\right] .
\end{multline}
Thus, the defect can be still characterized by the
transmission and reflection amplitudes of a free Majorana fermion.

In particular, we find that the transmission amplitude $T$
vanishes when
\begin{equation}
 \cos{\phi_1} + \cos{\phi_2} = 0 .
\end{equation}
This means that, even though the defect
coupling $\frakb$ is non-vanishing, the
low-energy universal properties are identical to the
open boundaries. 
Thus the open ends host Majorana zero modes in this case.
This is a rederivation of the findings in Ref.~\cite{KKWK2017}
for $t = \Delta$.

Furthermore, the present approach can be also extended to
the more general case of $t \neq \Delta$.
While the effective Hamiltonian in a naive continuum limit
is essentially the same whether $t=\Delta$ or not,
the structure of the eigenstates is different on the lattice
when $t \neq \Delta$.
For a given energy eigenvalue $\epsilon$,
the eigenequations on the lattice can be interpreted
as a recursion relation determining $\ttBphi_j$.
If and only if $t=\Delta$, $\ttBphi_{j+1}$ is completely
determined by $\ttBphi_j$ 
When $t \neq \Delta$, $\ttBphi_{j+1}$ is determined by
both $\ttBphi_{j-1}$ and $\ttBphi_j$.
This implies that, for a given energy eigenvalue $\epsilon$,
there are 4 linearly independent eigenstates.
Thus there must be 2 extra linearly independent eigenstates
in addition to the plane wave states $\ttBphi^+(p)$
and $\ttBphi^+(-p)$.
As we will see below, these extra states are bound state-like. 

First we consider the critical point $\mu=2t$ at zero energy.
The eigenequation in the bulk now reads
\begin{align}
\begin{pmatrix}
0 & t+\Delta
\\
t-\Delta & 0
\end{pmatrix} 
\ttBphi_{j-1}
- 2t
\begin{pmatrix}
 0 & 1 \\
 1  &  0
\end{pmatrix}
\ttBphi_j
+
\begin{pmatrix}
 0 & t-\Delta \\
 t+\Delta & 0
\end{pmatrix} 
\ttBphi_{j+1}
& =0.
\label{eq.latticeSch_Kzero}
\end{align}
This allows eigenstates
\begin{align}
 \ttBphi^L_j &= 
\begin{pmatrix}
 1 \\ 0
\end{pmatrix}
\left( \frac{t+\Delta}{t-\Delta}\right)^j,
\label{eq.ttBphiL}
\\
 \ttBphi^R_j &= 
\begin{pmatrix}
 0 \\ 1
\end{pmatrix}
\left( \frac{t-\Delta}{t+\Delta}\right)^j.
\label{eq.ttBphiR}
\end{align}
These represent eigenstates decaying to $j \to -\infty$ and
to $j \to \infty$, respectively.
Although we are interested in scattering of eigenstates
with finite momentum $p$ at a finite mass $m$ (and thus
at a finite energy $\epsilon$), corrections to the
eigenstates~\eqref{eq.ttBphiL},~\eqref{eq.ttBphiR}
are of $O(a)$ and negligible in the continuum limit.

Thus, including those states, the scattering states around the
defect may be given as
\begin{equation}
 \ttBphi_j =
\begin{dcases}
 \ttBphi^+(p) e^{i p a j}
+  R \ttBphi^+(-p) e^{- i p a j} 
+ c_L
\begin{pmatrix}
 1 \\ 0
\end{pmatrix}
\left( \frac{t+\Delta}{t-\Delta}\right)^j,
& ( j \leq 0),
\\
 T \ttBphi^+(p) e^{i p a j} 
+ c_R
\begin{pmatrix}
 1 \\ 0
\end{pmatrix}
\left( \frac{t+\Delta}{t-\Delta}\right)^{j-1},
& ( j \geq 1) ,
\end{dcases}
\end{equation}
where $c_L$ and $c_R$ are coefficients.

This scattering state must satisfy the matching conditions
at the defect. In the continuum limit near the critical point,
the matching conditions read 
\begin{align}
\frakb
\begin{pmatrix}
0 & t+\Delta
\\
t-\Delta & 0
\end{pmatrix} 
\ttBphi_0
- 2t
\begin{pmatrix}
 0 & 1 \\
 1  &  0
\end{pmatrix}
\ttBphi_1
+
\begin{pmatrix}
 0 & t-\Delta \\
 t+\Delta & 0
\end{pmatrix} 
\ttBphi_2
& =0.
\\
\frakb
\begin{pmatrix}
0 & t+\Delta
\\
t-\Delta & 0
\end{pmatrix} 
\ttBphi_{-1}
- 2t
\begin{pmatrix}
 0 & 1 \\
 1  &  0
\end{pmatrix}
\ttBphi_0
+
\frakb
\begin{pmatrix}
 0 & t-\Delta \\
 t+\Delta & 0
\end{pmatrix} 
\ttBphi_1
& =0.
\end{align}
Solving the matching conditions in terms of $c_L, c_R, R$, and $T$, 
the transmission amplitude $T$ is given as
\begin{equation}
T = \frac{1}{D} 
\left[
4 \frakb \Delta  p (\Delta +t) \left(\frakb^2 \Delta +\frakb^2 (-t)+\Delta +t\right) 
(\Delta  \cos{\phi_2} +t \cos{\phi_1} )
\right],
\end{equation}
where
\begin{multline}
D = \Delta^4 \left(\left(\frakb^2+1\right)^2 p+i
   \left(\frakb^2-1\right)^2 m\right)+\left(\frakb^2-1\right)^2 t^4 (p+i m)-4 i \left(\frakb^2-1\right)
   \Delta  t^3 (m-i p)
\\
+ 2 \frakb^2 \Delta ^2 
\left( \cos{2 \phi_2}
 \left(t^2 (p+i m)+2 \Delta  t (p+i m)+2 \Delta^2 p \right)
+t (p-i m) (4 \Delta  \cos{\phi_2} \cos{\phi_1}+t \cos{2 \phi_1})
\right)
\\
+2 \Delta^2 t^2 
\left(
 -\left(\frakb^4-3\right) p
 -i \left(\frakb^4+2 \frakb^2-3 \right) m
\right)+ 4 \Delta^3 t (p+i m) .
\end{multline}
We observe that the transmission amplitude $T$ vanishes when
\begin{equation}
 t \cos{\phi_1} + \Delta \cos{\phi_2} = 0.
\end{equation}
This again reproduces the Majorana zero modes around the defect
as pointed out in Ref.~\cite{KKWK2017}.

\bibliography{mybib}

\end{document}